\documentclass{aastex631}

\usepackage [english]{babel}
\usepackage [autostyle, english = american]{csquotes}
\MakeOuterQuote{"}
\newcommand{\degree}[1]{$^{\circ}$}
\newcommand{\mum}{$\mu$m\ }
\usepackage{url}
\shorttitle{Submillimeter-wavelength Polarimetry of IRC+10\,216}
\shortauthors{Andersson et al.}
\graphicspath{{./}{figures/}}

\begin{document}

\title{Submillimeter-wavelength Polarimetry of IRC+10\,216}

\author[0000-0001-6717-0686]{B-G Andersson}
\affiliation{Institute for Scientific Research, Boston College, 140 Commonwealth Avenue, Chestnut Hill, MA 02467, USA}
\affiliation{SOFIA Science Center, Universities Space Research Association, NASA Ames Research Center, M.S. N232-12, Moffett Field, CA 94035, USA}

\author[0000-0001-5996-3600]{Janik Karoly}
\affiliation{Jeremiah Horrocks Institute, University of Central Lancashire, Preston PR1 2HE, UK} 

\author[0000-0002-0794-3859]{Pierre Bastien}
\affiliation{Institut de Recherche sur les Exoplan\`etes (iREx) \& Centre de Recherche en Astrophysique du Qu\'ebec (CRAQ),\\ Universit\'e de Montr\'eal, D\'epartement de Physique, 
1375, Avenue Th\'er\`ese-Lavoie-Roux, Montr\'eal, QC, H2V 0B3, Canada}

\author[0000-0002-6386-2906]{Archana Soam}
\affiliation{Indian Institute of Astrophysics, II Block, Koramangala, Bengaluru 560034, India}
\affiliation{SOFIA Science Center, Universities Space Research Association, NASA Ames Research Center, M.S. N232-12, Moffett Field, CA 94035, USA}

\author[0000-0002-0859-0805]{Simon Coud\'e}
\affiliation{Department of Earth, Environment, and Physics, Worcester State University, Worcester, MA 01602, USA}
\affiliation{Center for Astrophysics $\vert$ Harvard \& Smithsonian, 60 Garden Street, Cambridge, MA 02138, USA}
\affiliation{SOFIA Science Center, Universities Space Research Association, NASA Ames Research Center, M.S. N232-12, Moffett Field, CA 94035, USA}

\author[0000-0001-8749-1436]{Mehrnoosh Tahani}
\affiliation{Banting and KIPAC Fellow: Kavli Institute for Particle Astrophysics \& Cosmology (KIPAC), Stanford University, Stanford, CA 94305, USA}
\affiliation{Covington Fellow: Dominion Radio Astrophysical Observatory, Herzberg Astronomy and Astrophysics Research Centre, National Research Council Canada, P. O. Box 248, Penticton, BC V2A 6J9 Canada}

\author[0000-0002-1913-2682]{Michael S.\ Gordon}
\affiliation{Ball Aerospace, 1600 Commerce St., Boulder, CO 80301}

\author{Sydney Fox-Middleton}
\affiliation{Physics Department, Santa Clara University, 500 El Camino Real, Santa Clara, CA 95053, USA}
\affiliation{SOFIA Science Center, Universities Space Research Association, NASA Ames Research Center, M.S. N232-12, Moffett Field, CA 94035, USA}



\begin{abstract}
We present SCUBA-2/POL-2 850 $\mu$m polarimetric observations of the circumstellar envelope (CSE) of the carbon-rich asymptotic giant branch (AGB) star IRC+10\,216.  Both FIR and optical polarization data indicate grains aligned with their long axis in the radial direction relative to the central star.  The 850 $\mu$m polarization does not show this simple structure.   The 850 $\mu$m data are indicative, albeit not conclusive, of a magnetic dipole geometry.  Assuming such a simple dipole geometry, the resulting 850 $\mu$m polarization geometry is consistent with both Zeeman observations and small-scale structure in the CSE.  While there is significant spectral line polarization contained within the SCUBA-2 850 $\mu$m pass-band for the source, it is unlikely that our broadband polarization results are dominated by line polarization.  To explain the required grain alignment, grain mineralogy effects, due to either fossil silicate grains from the earlier oxygen-rich AGB phase of the star, or due to the incorporation of ferromagnetic inclusions in the largest grains, may play a role.  We argue that the most likely explanation is due to a new alignment mechanism \citep{lazarian2020} wherein a charged grain, moving relative to the magnetic field, precesses around the induced electric field and therefore aligns with the magnetic field.  This mechanism is particularly attractive as the optical, FIR, and sub-mm wave polarization of the carbon dust can then be explained in a consistent way, differing simply due to the charge state of the grains.

\end{abstract}



\section{Introduction} \label{sec:intro}
Visual light interstellar medium (ISM) polarization was discovered in 1949 \citep{hiltner1949a,hiltner1949b,hall1949}, and from the start hypothesized to be associated with dichroic extinction by asymmetric dust grains aligned with the Galactic magnetic field.  The corresponding emission polarization was first detected by \citet{cudlip1982}.  Observations of dust-induced polarization covering the UV-to-sub-mm\footnote{For the purpose of the present paper, we will distinguish between Far Infrared (FIR) and sub-mm wave, as the former covering the wavelength range where the atmosphere is fully opaque from the ground, i.e. $\lambda\approx30-300\ \mu$m, and the latter being $\lambda\approx350\ \mu$m -- 1 mm.   In practice, here, the distinction is between SOFIA/HAWC+, and JCMT/SCUBA2/POL-2 or ALMA observations.} wavelength range serve as efficient probes of both the geometry and strength of the interstellar magnetic fields \citep[e.g.][]{davis1951b,chandrasekhar1953,houde2009,lazarian2022}, being widely observable with modest telescope size and observing time and straightforward to calibrate.  The general association of dust polarization with magnetic fields has been securely established, and the general mechanism of the grain alignment is now understood as being radiation driven \citep[cf][henceforth ALV15]{bga2015b}, but with many details still needing to be explored in detail.  

A quantitative theory of grain alignment based on radiative torques has been refined and tested over the last few decades \citep[e.g.][ALV15]{dolginov1976,draine1996,lazarian2007a}. This radiative alignment torque (RAT) theory predicts that the transfer of angular momentum from photons with wavelengths less than grain diameter ($\lambda<2a$, where \textit{a} is the effective grain radius) will spin up an irregular grain\footnote{The coupling to the radiation field does not fully cease at $\lambda>2a$, but falls with a steep dependence; \citep[cf][]{hoang2021}}.  For a paramagnetic grain, some of the rotational energy is exchanged for spin-flips in the solid \citep[the "Barnett effect";][]{purcell1979}.  This both causes the angular momentum of the grain to align with the grain's axis of maximum inertia, so-called "internal alignment", through Barnett dissipation  \citep[cf.][]{purcell1979,lazarian1999b} and, in steady state, magnetizes the grain. The magnetized grain then Larmor precesses around the external magnetic field ($B$-field), and continued radiative torques aligns the grains with the $B$-field (so called "B-RAT").  This paradigm has now been well-tested, including observations of a correlation with the radiation intensity \citep{medan2019,santos2019,soam2021a}, a correlation of aligned dust grain size with opacity into a cloud \citep{bga2007,vaillancourt2020}, a dependence on the angle between the radiation and the $B$-fields \citep{bga2011,vaillancourt2015}, and a loss of the grain alignment in the starless cores \citep[][ALV15]{jones2015,alves2014}.  However, several detailed aspects of the RAT paradigm are still in need of further clarification 

In particular, the role of grain mineralogy, and the influence of gas-grain relative velocities, remain to be fully understood.  With "mineralogy" we here refer to the composition and structure of the solid, as it influences particularly the magnetic properties (dia-, para- or ferromagnetism) of the grains.  One observational effect, related to the grain mineralogy, still needing to be fully understood is why carbonaceous grains appear not to be aligned in the interstellar medium (ISM).  

While the silicate feature at $\sim$9.7 $\mu$m shows significant polarization \citep{smith2000}, the aliphatic CH feature at 3.4 $\mu$m does not \citep{chiar2006}. As shown by \citet{chiar2013}, carbon grains will, when exposed to atomic hydrogen and UV radiation, develop CH bonds on their surfaces. The lack of polarization in the 3.4 $\mu$m feature therefore implies that carbonaceous grains are not aligned with any external reference direction in the ISM.  

\citet{chiar2006} also addressed the possibility, and behavior, of composite grains by comparing the fractional polarization ($p/\tau$), where p is the polarization in the line, and $\tau$ is the opacity in the (Stokes I) absorption line, in the silicate and CH features towards the Quintuplet cluster members GCS 3-II and 3-IV. They find that the upper limits on ($p/\tau)_{\mathrm{CH}}$ compared with the measured values for ($p/\tau)_{\mathrm{Sil}}$ argues against the model of composite grains with a silicate core and organic refractory mantles (ORM) and conclude that "\textit{the agent responsible for the hydrocarbon feature in the diffuse ISM is located in a grain population that is both physically separate from the silicates and far less efficient as a producer of polarization}".  

Recent theoretical work \citep[][]{draine2021} argues, however, that the observational constraints on the CH feature polarization are just above the expected level of polarization from composite grains consisting of a mix of amorphous silicates and other materials - so called "astrodust" - provided that the 3.4 $\mu$m absorption is preferentially located in grain surface layers.  Constraining the alignment characteristics of pure carbon dust therefore would be a valuable constraint on ISM modeling.

The non-detection of polarization in the 3.4 $\mu$m line is consistent with the FIR/sub-mm polarization spectrum models of \citet{draine2009}, which requires two dust components - a cooler aligned one and a warmer unaligned one.  Aligned silicate and unaligned carbon grains would explain the FIR/sub-mm wave polarization spectrum measured by \citet{vaillancourt2008}.  We note, however, that recent observations \citep[e.g.,][]{ashton2018} do not see the distinct concave spectrum discussed by \citet{vaillancourt2008}.  

As noted above, two kinds of alignment in dust grains are required to cause polarization.  "Internal alignment" is accomplished when the grain spins around one of its principal axes (usually the axis of maximal inertia = smallest axis; the lowest energy state for a given total angular momentum), yielding a constant projected shape of that grain.  Under angular momentum conservation, internal alignment requires efficient energy dissipation in the grain bulk, which \citet{purcell1979} showed was most efficiently accomplished through Barnett relaxation in paramagnetic grains \citep[cf][]{lazarian1999b}.  "External alignment" signifies the alignment of the individual grains' spin axes with an external reference direction, \textit{usually} the $B$-field.  Both types of alignment assume paramagnetic solids, which in the ISM applies to silicate grains.  Carbon grains are, however, diamagnetic and expected to show neither fast internal alignment nor to respond to magnetic fields.

Variants of RAT alignment are also possible, dependent on the radiation field strength and the grain mineralogy.  For grains with efficient internal alignment exposed to strong, highly anisotropic radiation fields, the external alignment direction is expected to change to the radiation field propagation direction (its "k-vector"; so called "k-RATs").  This phenomenon may have been detected in the Orion region \citep{chuss2019,pattle2021} and possibly in protostellar disks \citep{kataoka2017}.

RAT theory \textit{does} predict that also carbonaceous grains are spun up by an anisotropic radiation field (as seen in the laboratory by \citet{abbas2006}).  \citet{hoang2009a} have shown that for strong and anisotropic enough radiation fields, grains without internal alignment can experience a second-order k-RAT alignment.  Such alignment is likely to be weak and bi-modal, with some grains oriented with their minor axis along the radiation field direction and some with the major axis in that direction.  In regions with supersonic gas-grain drift, this drift can preferentially randomize the former grain-orientation, because of the larger cross-section towards the flow, leaving a net alignment of grains with their long axis in the gas-dust flow direction, thus causing polarization.

As discussed recently by \citet{lazarian2020}, a new alignment mechanism should exist for 1) charged carbon grains with 2) a net velocity perpendicular to the magnetic field.  Such grains get aligned via precession of their electric dipole moment around the induction electric field (\textbf{E} $\propto$ \textbf{v}$_{grain} \times$ \textbf{B}).  Both of the required conditions are likely fulfilled in the \textit{outer} envelope of AGB star winds, due to the radiative driving of the dust in the AGB wind, and the photoelectric grain charging by the UV light in the interstellar radiation field.  As noted by \citet{lazarian2020}, for inherently magnetically active grains (paramagnetic or stronger) B-RATs would, in most situations, dominate this mechanism.  Hence, such "E-RAT" alignment is observationally expected to be associated with carbon grains.  This electrically induced radiative alignment can cause the grains to align with either their long or short axis along the induced electric field, depending on the relative precession rate around the electrical and magnetic fields: ($\Omega_E$) and ($\Omega_B$).  This ratio is given by \citet[equation 9]{lazarian2020}

\begin{equation}
\frac{\Omega_E}{\Omega_B}=\frac{p_{el,J} V_{grain\perp}}{\mu c}
\end{equation}
\noindent
where p$_{el,J}$ is the grain's electric dipole moment parallel to its angular momentum, $V_{grain\perp}$ is the grain's velocity perpendicular to the direction of the magnetic field and $\mu$ is the grain's magnetic moment

As shown by \citet[][their Table 1, and Sec. 6.2]{lazarian2020}, if the precession rate around the E- or B-fields ($\Omega_E$) are faster than around the radiation field direction ($\Omega_k$), then, if  $\Omega_E > \Omega_B$, theory predicts that the observed emission polarization should be parallel to the $B$-field (opposite to the situation in interstellar - B-RAT - alignment).  If the precession rate around the B-field ($\Omega_B$) is faster, the polarization will be perpendicular to the $B$-field  (see \citet{lazarian2020} for details). Therefore, in addition to probing for the magnetic field geometry, such alignment would provide a measure of the micro-physics (including charging) of the grains. 

In the general ISM, silicate and carbonaceous dust are generally well mixed, making it very difficult to decisively probe this effect, even in environments, such as shock fronts, where systematic supersonic gas-dust drift may exist \citep{hoang2020}.  We have therefore initiated a program of observations targeting grain alignment in the chemically segregated environments of AGB star envelopes using FIR and sub-mm wave emission from the CSE, and optical polarimetry observations of background stars.  

As medium-mass stars first ascend the AGB their CSEs reflect the cosmic carbon-to-oxygen abundance ratio [C]/[O] of $<$1.  Because of the chemical stability of the CO molecule, all the carbon in these oxygen-rich CSEs is then tied up in that gas-phase molecule and the dust formed consists of silicates and metal oxide grains \citep{gail2013}.  As the thermal pulses of the later AGB evolution progress, \citep{lattanzio2004} newly synthesized carbon is dredged up to the stellar surface and ejected into the CSE, increasing the [C]/[O] ratio to unity (so called S-type stars) and eventually to $>$1 when the star becomes C-rich \citep{whittet2003,olofsson2004}.  The gas is now depleted of oxygen (which is tied up in CO), and the dust formed consists of various forms of carbon solids, including amorphous carbon and silicon carbide \citep[e.g.][]{ivezic1996a}.  

Observations of the planetary nebula BD\,+30$^\circ$3639 show that the size of the object at wavelengths dominated by silicate features is larger than that at wavelengths dominated by carbonaceous spectral features \citep{guzman-ramirez2015}, indicating that the O-to-C rich history of the shell may, in some sources, be traced through high-resolution observations.

IRC+10\,216 is a well known infrared object discovered in 1969 by Eric Becklin \citep{becklin1969} and identified as a dust enshrouded carbon star the following year \citep{miller1970}. At a distance of only $d=123\pm$14\ pc; \citep{groenewegen2012} it can be resolved by many telescopes, and it has been extensively studied both in continuum and line radiation \citep[e.g.][and refs. therein]{cernicharo2010,decin2011}.  The extensive dust and gas circumstellar shell ends in a termination shock, where the CSE runs into the ISM, seen in both the ultraviolet \citep{sahai2010} and the FIR \citep{ladjal2010}.  
The effective extent of the CSE of IRC+10\,216 depends on the wavelength of observation and its depth.  For reference, the CSE can be detected in visible light (azimuthally averaged to $\sim$200\arcsec \citep{mauron2000}, the 70$\mu$m extended emission can be detected to at least 285\arcsec, \citep{dharmawardena2018} and the asymmetric astrosphere is located at $\sim$ 500-600\arcsec \citep{sahai2010} from the central star.

With a high C/O ratio (C/O=1.4 \citet{winters1994}, \citet[][and ref.s therein]{milam2009}), the dust is fully carbonaceous, consisting of 95\% amorphous carbon and 5\% SiC by mass; \citet{ivezic1996a}.  While iron is found to be significantly depleted in the gas phase of the CSE \citep{mauron2010}, no direct observational evidence for iron grains or iron inclusions in the carbon grains, has been found.

Evidence for magnetic fields in the IRC+10\,216 CSE come from observations of the Goldreich-Kylafis (G-K) effect in several molecules \citep{girart2012} with the Submillimeter Array (SMA), and from Zeeman observation in the J=1-0 line of CN \citep{duthu2017} with the IRAM 30 m telescope.  The G-K observations on small scales indicate a "global radial pattern" \citep{girart2012}.  However, the CN Zeeman observation shows significant spatial variations (including field direction reversals) in the line of sight field strength around the star, indicating deviations from spherical symmetry.  Hence, while there are strong indications of a magnetic field in the CSE, the geometry of the field is not fully clear. Evidence for simple dipole geometries in AGB star CSEs come, for instance, from \citet{szymczak2001}, using OH maser emission and \citet{vlemmings2005}, using H$_2$O masers.  Both find that the magnetic field on hundreds-to-thousands of AUs scales in the AGB star VX Sgr \citep{tabernero2021} can be modeled as resulting from a dipole.

For sources such as AGB stars, observed with unresolved aperture or centered slit polarimetry, asymmetries in the dust distribution and grain alignment effects are very difficult to separate.  \citet{kahane1997,bastien2003} acquired spatially unresolved aperture polarimetry of a sample of 68 AGB stars in a band centered at 0.883$\mu$m.  They divided their sample into two groups, spherical and aspherical envelopes, based on their CO line profiles. Both polarization and CO line profiles can reveal the presence of non-spherical envelopes.  No statistical difference was found in polarization histograms of stars with “normal” (spherical) and “abnormal” (aspherical) CO line profiles \citep{kahane1997}. However, the polarization data showed a significant difference between C-rich and O-rich envelopes, with C-rich CSE presenting a higher polarization than their O-rich counterparts \citep{kahane1997,bastien2003}, indicating a more asymmetric dust distribution in the former.  \citet{bieging2006} performed optical (0.42 - 0.84$\mu$m) spectro-polarimetry of 21 AGB stars, 13 proto-planetary nebulae, and two R CrB-type stars. They, also, found a higher fraction of polarization for the carbon rich AGB stars (five of six) than for the oxygen rich stars (eight of fourteen), which they attributed to the later developmental stage of the carbon-rich stars.  All of these observations are, however, likely dominated by scattering polarization.
High-resolution, Active Optics (AO) supported, observations in the optical and near infrared can resolve the inner part of the CSEs \citep[e.g.][]{kastner1994,kastner1996,jeffers2014,Khouri2020,montarges2023} and provide important information about the spatial dust distribution and grain size distribution. However, since these observations are dominated by [Rayleigh] scattering, they do not provide information about the grain alignment.  In most radiative transfer modeling of these kinds of data, \citep[e.g.][]{montarges2023} the grains are either assumed to be symmetrical or have a random spatial orientation.
However, imaging FIR and sub-mm wave observations can resolve the CSEs and can differentiate between the two mechanisms of polarization (scattering and dichroic emission), as the emission wavelengths are much larger than the grain sizes. Similarly, optical polarimetry of background stars with careful subtraction of the diffuse, scattered, light from the CSE can show how the grains are aligned. 

IRC+10\,216 has been previously observed at multiple wavelengths in scattering, dichroic extinction, and emission polarization.  Optical through near-infrared polarization observation of the central part of the CSE has been reported by e.g. \citet{dyck1971, cohen1982, trammell1994} showing significant polarization rising to the blue (40\% at 0.67 $\mu$m \citep{cohen1982}), consistent with Rayleigh (dust) scattering.  The small scale dust distribution (2-3\arcsec\ scales, and hence much smaller than the beams in our studies) has also been extensively studied in recent years. \citet{kastner1994,kastner1996} mapped the large-scale structure of the CSE at J, H and K-bands, while \citet{murakawa2005}, using 2 $\mu$m polarization, found a NW-SE structure that they identified as a possible dust torus.  Optical broad-band polarimetry \citep{jeffers2014} reveal a similar structure, which is also consistent with multi-epoch HST imaging \citep{kim2021}.  In an accompanying paper \citep[Andersson et al., 2023b, in preparation, ][]{bga2018}\footnote{See also presentation no. 5, e-proceedings of "From Stars to Galaxies II": \url{http://cosmicorigins.space/fstgii}}  we discuss polarimetry of background stars shining through the CSE.  While we detect strong scattered light polarization, we can, with careful sky-subtraction, also determine the dichroic extinction polarization towards these stars. 

Our SOFIA/HAWC+ observations at 53 $\mu$m \citep[at 5\arcsec\ resolution][]{bga2022a} show a uniform radial polarization for IRC+10\,216.  The FIR polarization maps at 154 and 214 $\mu$m also show predominantly radial polarization, albeit with lower S/N and at worse resolution.  A tight correlation between the FIR polarization and dust temperature indicates radiative grain alignment, via the second-order k-RAT process discussed above.  The low polarization efficiency seen in the inner parts of the 850 $\mu$m data presented here (Figure \ref{Fig:p_vs_e}) is consistent with k-RAT alignment of grains with no (or slow) internal alignment \citep{hoang2009a}.

Here we present new JCMT/SCUBA-2/POL-2 observations of the IRC+10\,216 CSE and analyze these in the context of SOFIA/HAWC+ and optical polarimetry (as well as other archival data sets) and the general RAT grain alignment paradigm. Along with \citet{bga2022a}, this paper is the start of a comprehensive study we've named the Survey of Polarization in AGB Circumstellar Envelopes (SPACE). We are working to complement the studies of IRC+10216 with observations of O-rich and S-type AGB stars IK Tau and W Aql. 

The paper is structured as follows: We present the observations and data reduction method in Section~\ref{sec:obs}. Section~\ref{sec:analysis} introduces the analysis of our results and Section~\ref{sec:res} summarizes these before Section~\ref{sec:Disc} where we discuss our results and possible alignment mechanisms. Finally, we summarize our findings in Section~\ref{sec:concl}.

\section{Observations and Data Reduction} \label{sec:obs}

IRC+10\,216 was observed with the SCUBA-2/POL-2 instrument combination \citep{holland2013,bastien2011} on the James Clark Maxwell Telescope, during 2018 January 11 and 12.  Six observations lasting 41 minutes each of IRC+10\,216 were carried out.  The regular POL-CV daisy pattern for SCUBA-2/POL-2 covers the star and all of its circumstellar envelope well.  We acquired 12 such patterns to reach a sensitivity of about 3 mJy/beam in polarized intensity in 8 arcsec$^2$ pixels.  Six additional observations were acquired in flexible mode on January 19. We have also included 8 repeats that are $\sim$31.5 minutes long from archival observations on CADC\footnote{Canadian Astronomy Data Centre: \url{https://www.cadc-ccda.hia-iha.nrc-cnrc.gc.ca/en/}} (Project ID: M19BP001; PI: Peter Scicluna).

During one observation on January 11, the telescope reached an elevation of 83\arcdeg. However, upon inspection of the data from that observation, no distortions of the data due to the JCMT alt-az mount could be detected.
The membrane in front of the telescope, in regular observations, was removed during the period 2017 December 5 to 2018 January 10 to carry out commissioning tests targeted at better characterizing the instrumental polarization due to the telescope and membrane. The data for this project were, therefore, acquired just after the reinstallation of the Gore-Tex membrane in front of the telescope.   Comparison of the instrumental polarization after the membrane was reinstalled with those before it was removed showed no significant differences. 

IRC+10\,216 was previously observed with the earlier JCMT sub-mm wave polarimeter SCUPOL.  These observations were reprocessed, combined and presented by \citet{matthews2009}.  Because these data were of lower signal-to-noise, we have not included them in the present analysis.

We reduced the JCMT/POL-2 850 $\mu$m observations using the {\tt\string pol2map} command of the SMURF package \citep{chapin2011} in the Starlink software \citep{currie2014}. In the first step, using {\tt\string pol2map}, the Stokes~\textit{I}, \textit{Q} and \textit{U} time streams are separated from the raw observations using the {\tt\string calcqu} command. Then the Stokes \textit{I} time streams are processed using the command {\tt\string makemap} to produce an initial Stokes~\textit{I} map. This Stokes~\textit{I} map is used as a fixed-SNR mask for the second step of the reduction where final Stokes~\textit{I}, \textit{Q} and \textit{U} maps are made. The second step consists of running {\tt\string pol2map} again, however we use {\tt\string skyloop}\footnote{\url{http://starlink.eao.hawaii.edu/docs/sc22.pdf}} to make the final maps rather than {\tt\string makemap}. {\tt\string Skyloop} is an iterative map-making command that reduces the growth of large-scale structures that may form due to the map-making routine by combining all the observations at each iteration, rather than each observation being created individually and then combined at the end. A polarization vector catalog is then created using the Stokes~\textit{I}, \textit{Q} and \textit{U} maps. The data were reduced using a pixel size of 8$\arcsec$ with a 14$\arcsec$ beam. The final Stokes~\textit{I}, \textit{Q}, and \textit{U} maps were flux calibrated, in units of $\rm mJy\,beam^{-1}$, using a flux calibration factor (FCF) for 850\,$\mu$m of 748 $\rm Jy\, beam^{-1}\, pW^{-1}$ and in units of $\rm mJy\,arcsec^{-2}$ using a FCF of 2.93 $\rm Jy\, arcsec^{-2}\, pW^{-1}$ \footnote{These conversions were done using the CALIBRATE-SCUBA-2-DATA recipe under the PICARD package in STARLINK.}.

The final polarization values in the vector catalog are debiased using the Stokes \textit{Q} and \textit{U} variances to remove the statistical bias in regions of low signal-to-noise ratio \citep[SNR;][]{wardle1974}. 

The values for the debiased polarization $P$ were calculated from 

\begin{equation}
P=\frac{1}{I}\sqrt{Q^{2}+U^{2}-\frac{1}{2}(\delta Q^{2}+\delta U^{2})}   \,\,,
\end{equation}
where \textit{I}, \textit{Q}, and \textit{U} are the Stokes parameters, and $\delta Q$, and $\delta U$ are the uncertainties for Stokes \textit{Q} and \textit{U}. The uncertainty $\delta P$ of polarization was obtained using

\begin{equation}
\delta P = \sqrt{\frac{(Q^2\delta Q^2 + U^2\delta U^2)}{I^2(Q^2+U^2)} + \frac{\delta I^2(Q^2+U^2)}{I^4}}  \,\,,
\end{equation}
with $\delta I$ being the uncertainty for the Stokes~\textit{I} total intensity. 

The polarization position angles $\theta$, increasing from north to east in the sky projection, were measured using the relation 

\begin{equation}
{\theta = \frac{1}{2}\tan^{-1}\frac{U}{Q}} \,.
\label{eq:polangle}
\end{equation}

The corresponding uncertainties in $\theta$ were calculated using

\begin{equation}
\delta\theta = \frac{1}{2}\frac{\sqrt{Q^2\delta U^2+ U^2\delta Q^2}}{(Q^2+U^2)} \times\frac{180^{\circ}}{\pi}  \,\, 
\label{eq:dtheta}
\end{equation}

Vectors were then selected using SNR cuts of $I >$ 20 $\delta I$ and $P >$ 2.5 $\delta P$. The selection criteria were chosen as a compromise between high SNR constraints and maintaining a significant number of polarization vectors. The cut in polarization is roughly equal to an uncertainty in the position angle of $\delta\theta = 11.5^{\circ}$ \citep{naghizadeh1993} which is quite reasonable. The SNR constraints are common throughout studies using SCUBA-2/POL-2 data \citep[e.g.][]{kwon2018,liu2019b}.

\begin{figure}
	\centering
	\resizebox{12 cm}{10cm}{\includegraphics{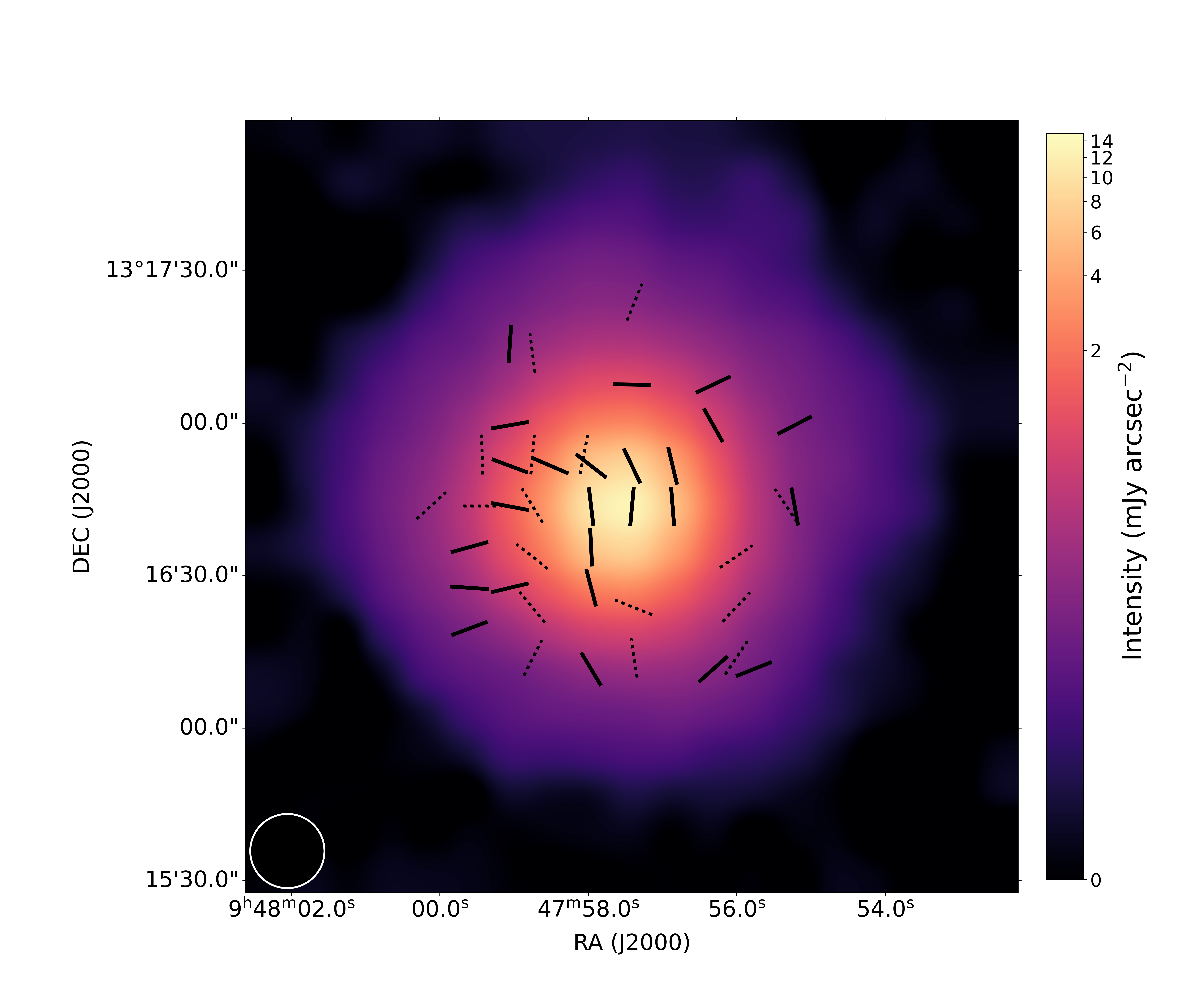}}
	\caption{The SCUPOL polarization vectors \citep{matthews2009} with $I/\delta_I>0$ and $P/\delta_P>2$ are shown as black dotted vectors (binned to a 10$\arcsec$ grid), with our POL-2 polarization vectors in solid black (on an 8$\arcsec$ grid) with $I/\delta_I>20$ and $P/\delta_P>2.5$. The POL-2 data has better sensitivity than SCUPOL data, with approximately 50\% more vectors at the same SNR cut. The color background shows the POL-2 Stokes \textit{I} component with the SCUBA-2/POL-2 850~$\mu$m beam size shown in the bottom left.  All beam sizes quoted in this paper are in terms of full-width at half maximum (FWHM)} 
	\label{SCUPol_Pol2_fig}
\end{figure}

The earlier observations of IRC+10\,216 with SCUPOL \citep{matthews2009} are of lower SNR with cuts of $I >$ 0 and $P >$ 2 $\delta P$. We show the SCUPOL polarization vectors plotted with our POL-2 vectors in Figure~\ref{SCUPol_Pol2_fig}. When applying a similar SNR cut to the SCUPOL vectors as our own, we have $\sim$50\% more vectors with POL-2, increasing from 15 to 22 vectors. We can also compare the vectors that spatially match. There are only seven vector pairs which spatially match within 8$\arcsec$ which we have chosen due to our vector catalog being binned to 8$\arcsec$. While the polarization amount is generally mutually consistent, the polarization angles vary significantly, with only three vectors agreeing well within the position angle uncertainties for moderate SNR data \citep{naghizadeh1993}. We note that SCUPOL was a different polarimeter and observed in a different mode than POL-2 which is a significant upgrade in sensitivity and ability to deal with atmospheric variations. So we do not expect perfect agreement, but the POL-2 data are of higher quality.

\begin{figure}
	\centering
         \resizebox{18cm}{5.5cm}{\includegraphics{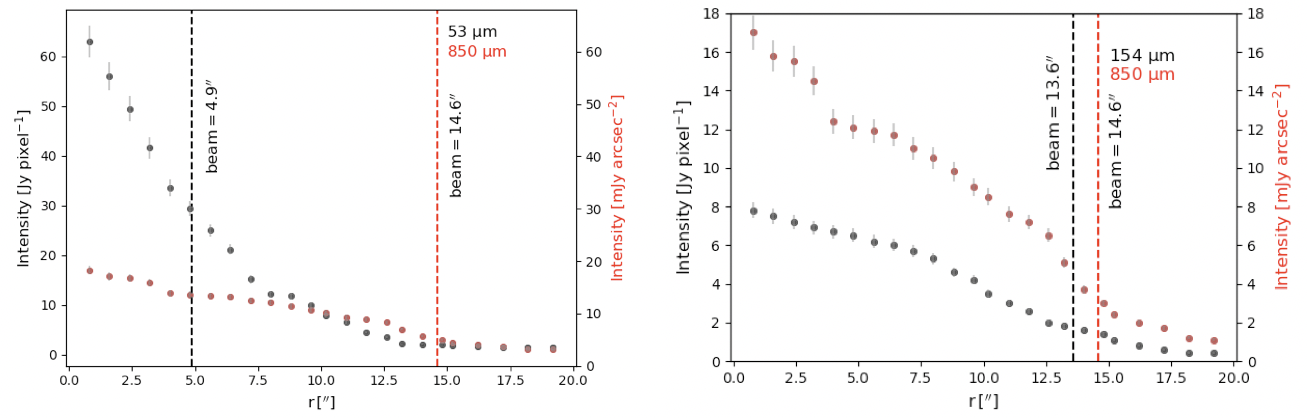}}
	\caption{(Left panel) The variation of Stokes $I$ intensity with the projected radius of IRC+10\,216 in 53 and 850 $\mu$m wavelengths is shown. (These are azimuthally averaged fluxes around the star. The shells of 2.5$\arcsec$ widths were drawn around the central star and average flux was extracted in those shells.)  The right panel shows the same plot, but for 154 and 850 $\mu$m wavelengths. Vertical dashed lines indicate the beam sizes in respective wavelengths.} \label{Fig:I_vs_r}
\end{figure}

Figure \ref{Fig:I_vs_r} shows the azimuthally averaged variation of intensities of different wavelength emissions with the offset from  IRC+10\,216. The beam-sizes corresponding to each observation are also indicated in the plots. The variation of 53 $\mu$m intensity with radius is clearly different from that of 850 $\mu$m, while both are decreasing with offset. However, a similarity in the intensity variations of 154 and 850 $\mu$m can be noticed in the right panel of the figure.  The data for the 53 and 154 $\mu$m traces are taken from \citet{bga2022a}.

\section{Analysis}
\label{sec:analysis}

Both at small \citep[e.g.][]{debeck2012} and large \citep[e.g.][]{mauron1999} scales, the over-all density of the shell is well-described by a $n \propto r^{-2}$ density, out to the astrosphere \citep{sahai2010,ladjal2010}.  Modeling using both DUSTY \citep{ivezic1999} e.g. \citet{ivezic1996a,bga2022a} and MCMax \citep{min2009}, e.g. \citet{debeck2012} yield good agreement between the emitted spectra and a monotonously decreasing temperature distribution in the envelope \citep[e.g.][]{ivezic1996b,debeck2012,bga2022a}. The emission at both FIR and sub-mm wavelengths are optically thin \citep[][Table 4 of the latter]{ivezic1996a,bga2022a}, and for the radial offsets from the star considered here, the dust temperature is low enough ($\lesssim$200 K) that all our observed wavelengths are well into the Rayleigh-Jean limit.  Therefore, the Stokes $I$ emission from the shell, for all wavelengths considered, is dominated by the density profile in the CSE.

\subsection{Radial coordinate polarization}\label{sec:rad_pol}

To better study the radial polarization in IRC+10\,216, we use "radial Stokes parameters" $Q_r$ and $U_r$, as defined in equation~\ref{eq:radialStokes}, where the reference frame is rotated by the polar angle at each point with respect to the star location:
\begin{equation}
Q_r = +Q\cos{2\xi} +U\sin{2\xi}, ~~
U_r = -Q\sin2\xi + U\cos 2\xi, 
\label{eq:radialStokes}
\end{equation}
where $\xi = \arctan{\frac{x-x_0}{y-y_0}}$ is the polar angle of a given pixel coordinate position $(x,y)$ (corresponding to equatorial coordinates) and $(x_0, y_0)$ is the location of the star. Because of the moderate Declination of IRC+10\,216 ($\sim$13.25$^\circ$) we have omitted the cos($\delta$) factor for the Right Ascension distance in the numerator.

Positive $Q_r$ shows radial polarization, while negative $Q_r$ indicates tangential polarization. $U_r$ represents polarization with an angle of $\pm 45^{\circ}$ from the radial direction. Therefore, zero $U_r$ indicates purely radial or tangential polarization.

\begin{figure}
\centering
\includegraphics[scale=0.43, trim={0cm 0cm 0cm 0cm},clip]{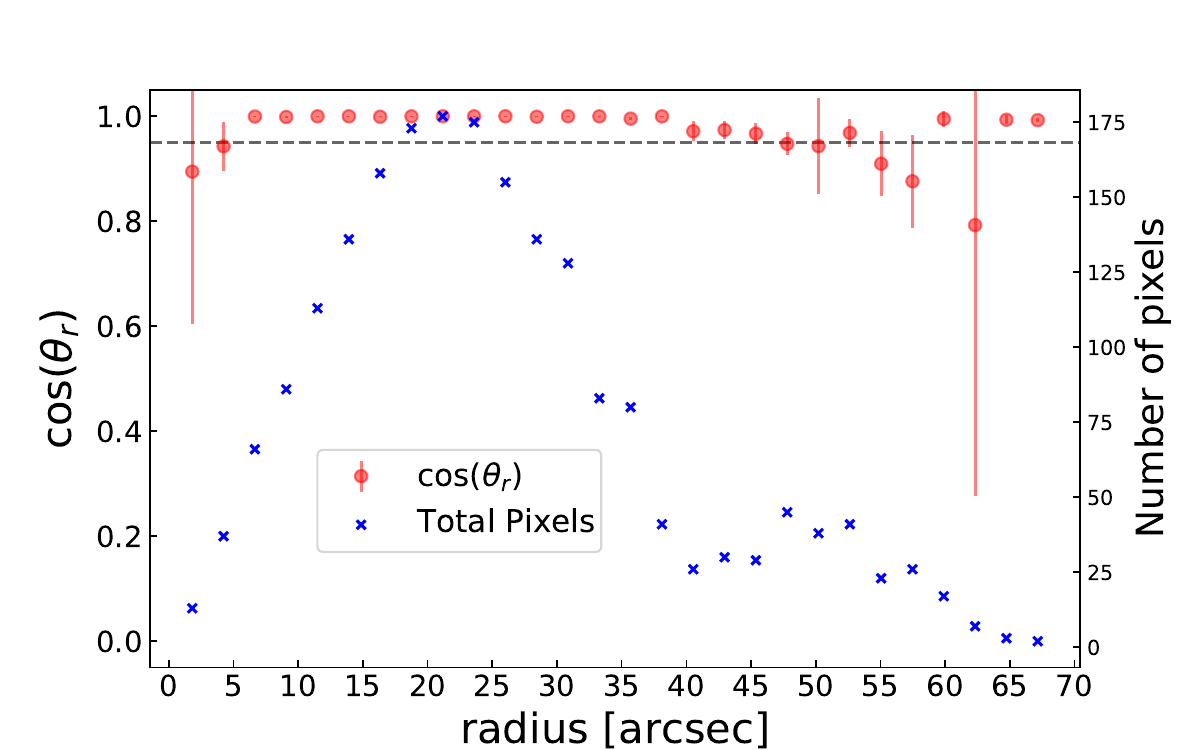}
\includegraphics[scale=0.43, trim={0cm 0cm 0cm 0cm},clip]{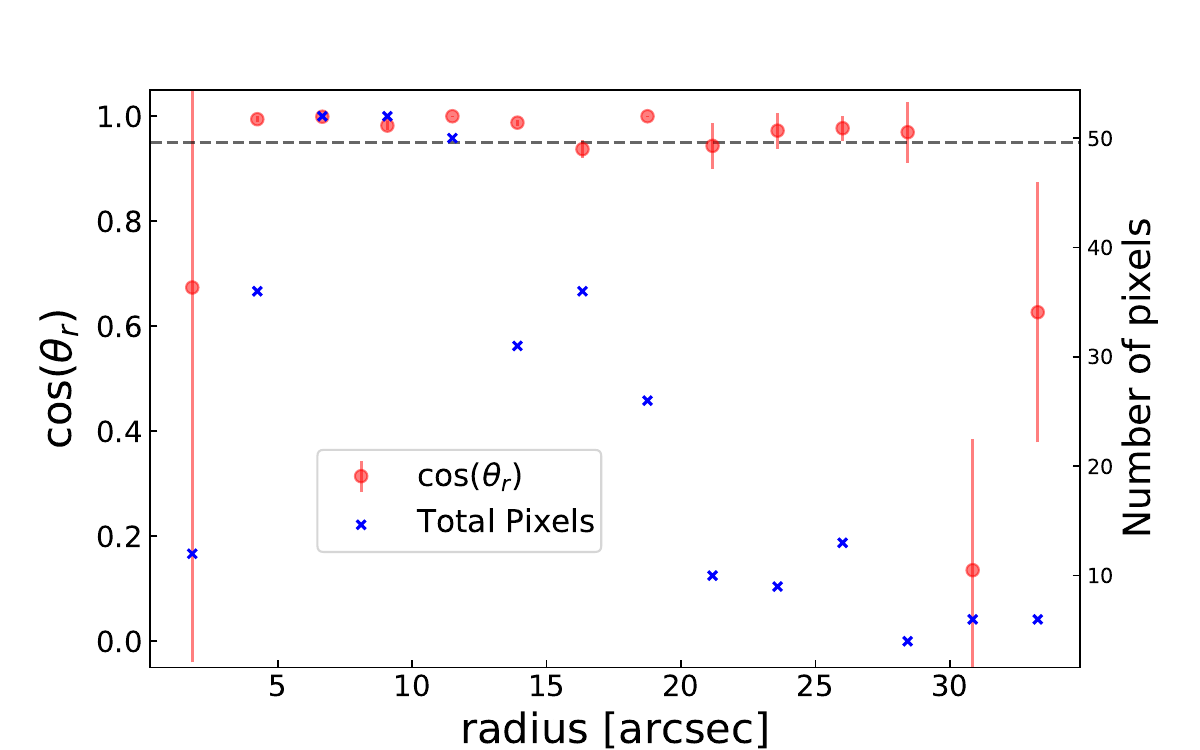}\\
\includegraphics[scale=0.43, trim={0cm 0cm 0cm 0cm},clip]{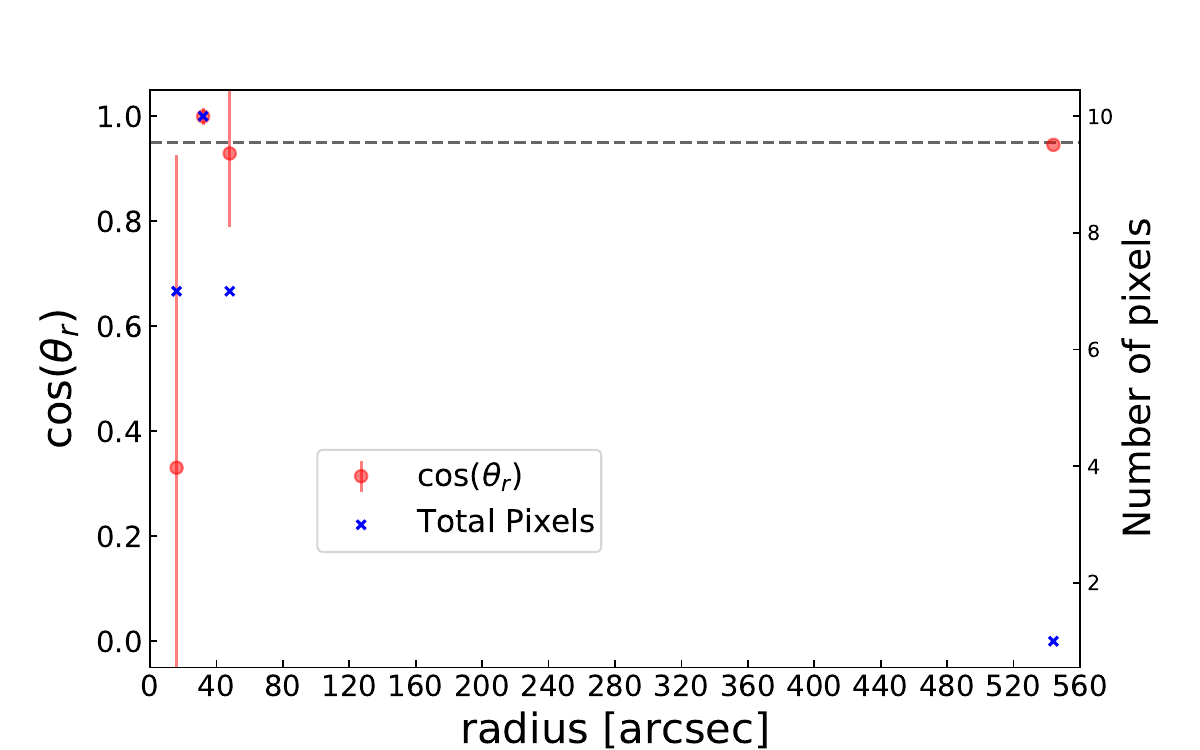}
\includegraphics[scale=0.58, trim={0cm 0cm 0cm 0cm},clip]{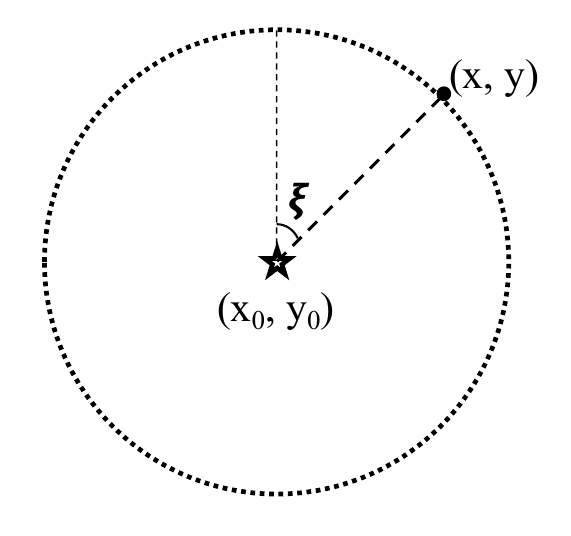}
\caption{Radial polarization with SNR($I$)$>20$ and SNR($P$)$>2.5$. The $x$ and $y$ axes represent the radial profile of $\cos{\theta_r}$ as described in equation~\ref{eq:theta} and the radius from the center of the star. The dashed horizontal line shows $\cos{\theta_r} = 0.95$, with points above this line (including their error bars) representing radial polarization (allowing for 5\% uncertainty; see \citet{tahani2023} for more detail). The blue dots illustrate the number of pixels (satisfying the SNR cut) in each annulus.  Upper left panel: HAWC$+$ 53\,$\mu$m observations. Upper right panel: HAWC$+$ 154\,$\mu$m observations. Lower left panel: JCMT POL-2 observations. Lower right panel: A cartoon illustrating the radial polarization frame (see Eq. \ref{eq:radialStokes}).} 
\label{fig:Radial}
\end{figure}

We take thin annuli ($\sim 2.2\arcsec$ in HAWC$+$ and $\sim 14\arcsec$ in JCMT observations) around the star and find the averaged $Q_r$ and $U_r$ in each annulus to identify the radial profile of $Q_r$ and $U_r$. Following that we determine 
\begin{equation}
   \theta_r = 0.5 \times \arctan\big(\frac{<U_r>}{ <Q_r>}\big), 
   \label{eq:theta}
\end{equation}
as defined by \citet{tahani2023} and shown in Figure~\ref{fig:Radial} for the HAWC$+$ 53\,$\mu$m (upper left panel), HAWC$+$ 154\,$\mu$m (upper right panel) and the JCMT POL-2 850\,$\mu$m observations (lower panel). The data have the same constraints of SNR($I$)$>20$ and SNR($P$)$>2.5$. Allowing for 5\% uncertainty, a point with $\cos{\theta_r} > 0.95$ (including their error bar) represents radial polarization at that radius.  This choice is clearly somewhat arbitrary, but as can be seen from Figure~\ref{fig:Radial} provides a reasonable separation between systematically radial and non-radial polarization.  The HAWC$+$ 53\,$\mu$m observations show clear radial polarization between 6\arcsec and 45\arcsec, while the JCMT observations do not indicate a clear radial polarization. The HAWC$+$ 154\,$\mu$m observations show radial polarization only from 5\arcsec to 28\arcsec.

The blue x markers in Figure \ref{fig:Radial} show the number of pixels with polarization measurements included for each radial bin.  Comparing the uncertainties indicated for the cos($\theta_r$) values with the variation in the number of points per bin shows that the dominant source for the larger uncertainties is dispersion within each sample, rather than localized higher measurement errors.

\subsection{CN Zeeman}

\begin{figure}[h]
	\centering
	\resizebox{9cm}{6cm}{\includegraphics{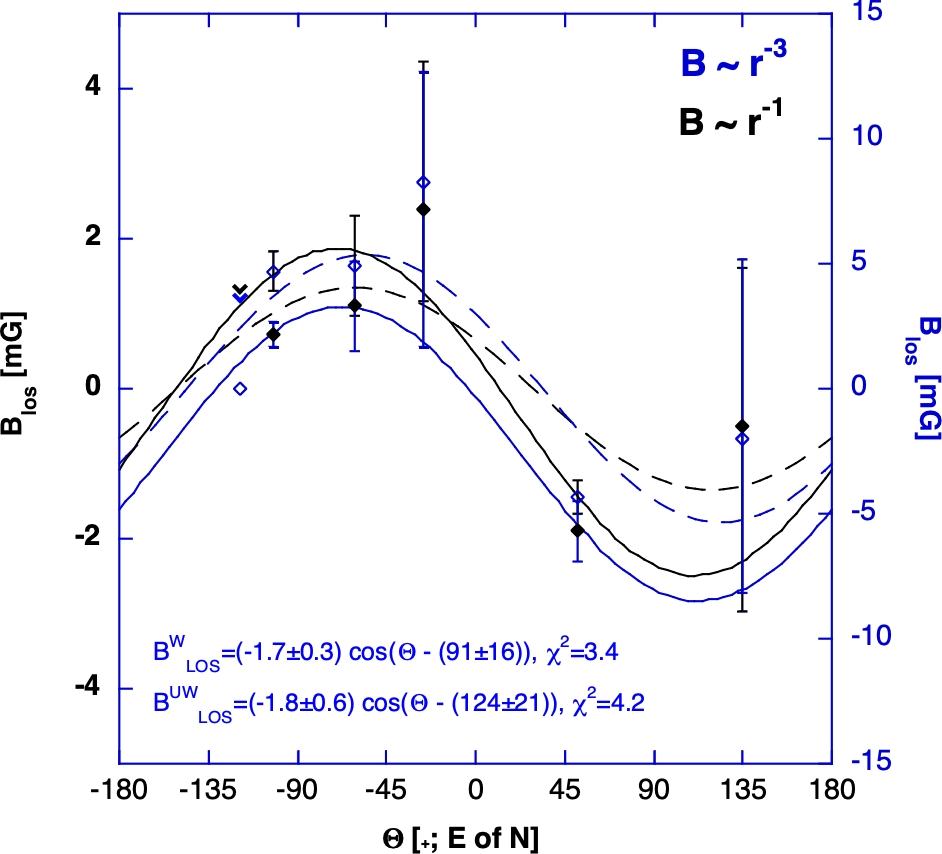}}
	\caption{The line-of-sight magnetic field strength, based on the CN (J=1-0) Zeeman observations of \citet{duthu2017}, scaled to a common radial distance from the central star using an r$^{-3}$ scaling (blue open symbols and curves, right axis) and using an r$^{-1}$ scaling (black, filled, symbols and curves, left axis).  The weighted (W; solid lines) and unweighted (UW; dashed lines) fits yield consistent results, both in amplitude and phase (see text for details).}
	\label{Fig:zeeman_plot}
\end{figure}

\citet{duthu2017} observed the Zeeman effect in the IRC+10\,216 CSE, using the CN (J=1–0) line at 113 GHz with the IRAM 30m telescope. They detected the line-of-sight magnetic field in five locations around the central star (plus one upper limit) with radial offsets ranging from 18 to 27$\arcsec$.  Because of the varying distances from the star, they - and we - scale the observations to a common distance from the star to allow comparisons with models.  They normalized their measurements to a common distance using a $r^{-1}$ dependence, based on \citet{vlemmings2012}.  As shown by Figure \ref{Fig:zeeman_plot} a coherent variation of the line of sight $B$-field is indicated by these data. We used this $r^{-1}$ scaling, as well as a $r^{-3}$ scaling, appropriate for a dipole field, and fit the resulting values to an azimuthal cosine function.  
\begin{equation}
B_{LOS}=a\cdot cos(\Theta - \Theta_0)
\end{equation}
With the limited number of data points, and marginal signal-to-noise these fits are not conclusive, but provide an indication of the magnetic field structure in the CSE, especially the orientation of a possible dipole field orientation.  Figure \ref{Fig:zeeman_plot} shows the best fit cosine functions (weighted, in solid lines, and unweighted, in dashed lines) of the \citet{duthu2017} data for both scalings ($r^{-3}$ in blue and $r^{-1}$ in black).  We show both types of fits, since we scaled also the uncertainties in the $B$-field by the two radial scaling.  In the fits, we have set the value at (-18, -10) - an upper limit in \citet{duthu2017} - to zero and assigned the upper limit as the uncertainty (indicated in Figure \ref{Fig:zeeman_plot} by wedges showing the upper limit and an open symbol for the value used in the fits).  Consistent results are found if the point is simply omitted.  The various fits give mutually consistent results.  In all cases, a pole orientation around 90-125$^\circ$ East of North is found, albeit with significant uncertainties.  For the $r^{-1}$ scaling, the weighted and unweighted fits, we find pole-orientations of 113$\pm$17$^\circ$ and 119$\pm$21$^\circ$, respectively.  For the $r^{-3}$ scaling, the weighted and unweighted fits yield pole-orientations of 91$\pm$16$^\circ$ and 124$\pm$21$^\circ$, respectively.  

\subsection{Dipole Modeling}

Given the clearly non-radial polarization pattern in the 850 $\mu$m data, contrasting with the FIR HAWC+ 53 and 154 $\mu$m results, an alternative polarization geometry is required - unless the geometry is inherently random.  However, a truly random orientation of the grains and/or geometry, would produce a zero polarization.  Hence, we need an ordered structure in the CSE which could allow grain alignment and produce the non-radial polarization pattern observed.  Based on the results of \citet{szymczak2001} and \citet{vlemmings2005}, the fits to the Zeeman data discussed above, and the simplicity of geometry (lowest order terms of a spherical harmonic expansion), we chose this to be a projected dipole field.  This implies a \textit{magnetic} dipole, but does not at this point require it.

We therefore compared the observed 850 $\mu$m polarization pattern to a simple model of a projected magnetic dipole of the form described in Appendix \ref{A1}.  Because of the simplifying assumptions in our model, we only use the position angles, but not the amount of polarization, in our fitting.  We generated a family of models with varying rotation angles $\Theta$, where $\Theta$ is the rotation of the dipole model (E-vectors) in the plane-of-sky (POS).  We took the CSE to be the central area where $I/\delta_I>10$ (though we maintained our $I/\delta_I>20$ cut on polarization vectors in this area) which was a region with diameter of $\approx$104$\arcsec$. We generated the dipole on a 104$\times$104 grid and set the spacing to 1$\arcsec$. We sought a maximum likelihood estimate (MLE) \citep{bevington1992} of the model rotation angle, by performing a $\chi^2_{\rm red}$ minimization of the model fits to the polarization position angle data. We used the $\chi^2_{\rm red}$ normalization in the least-squares MLE since this provides an indication of the absolute goodness of the model fits.  Because the uncertainties for polarization data (amount, $p$ and position angle $\theta$) for marginal signal-to-noise are not Gaussian, care has to be taken in such minimization.  We used the analysis of such non-Gaussian position angle uncertainty distributions by \citealt[][specifically their eq. 3 \& 4]{naghizadeh1993} to assign the weights in the $\chi^2_{\rm red}$ modeling of the position angles.  With our limited selection of only vectors with p $>$ 2.5$\delta_{p}$ this should be a minor concern but we include the factor of 1.2 in Eq.~\ref{eq:chi}. 

\begin{figure*}
	\centering
        \includegraphics[scale=0.2,angle=0]{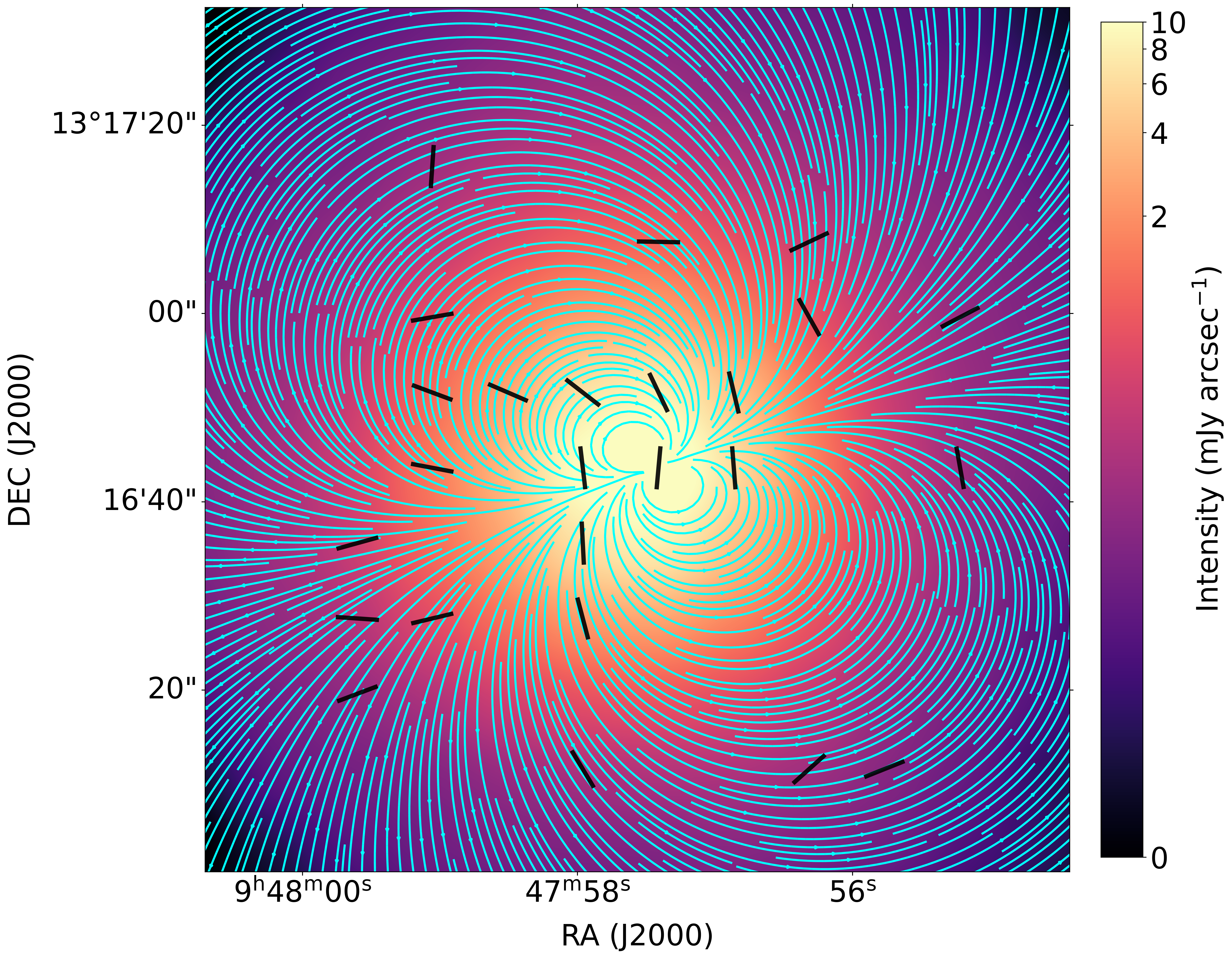}
        \includegraphics[scale=0.2,angle=0]{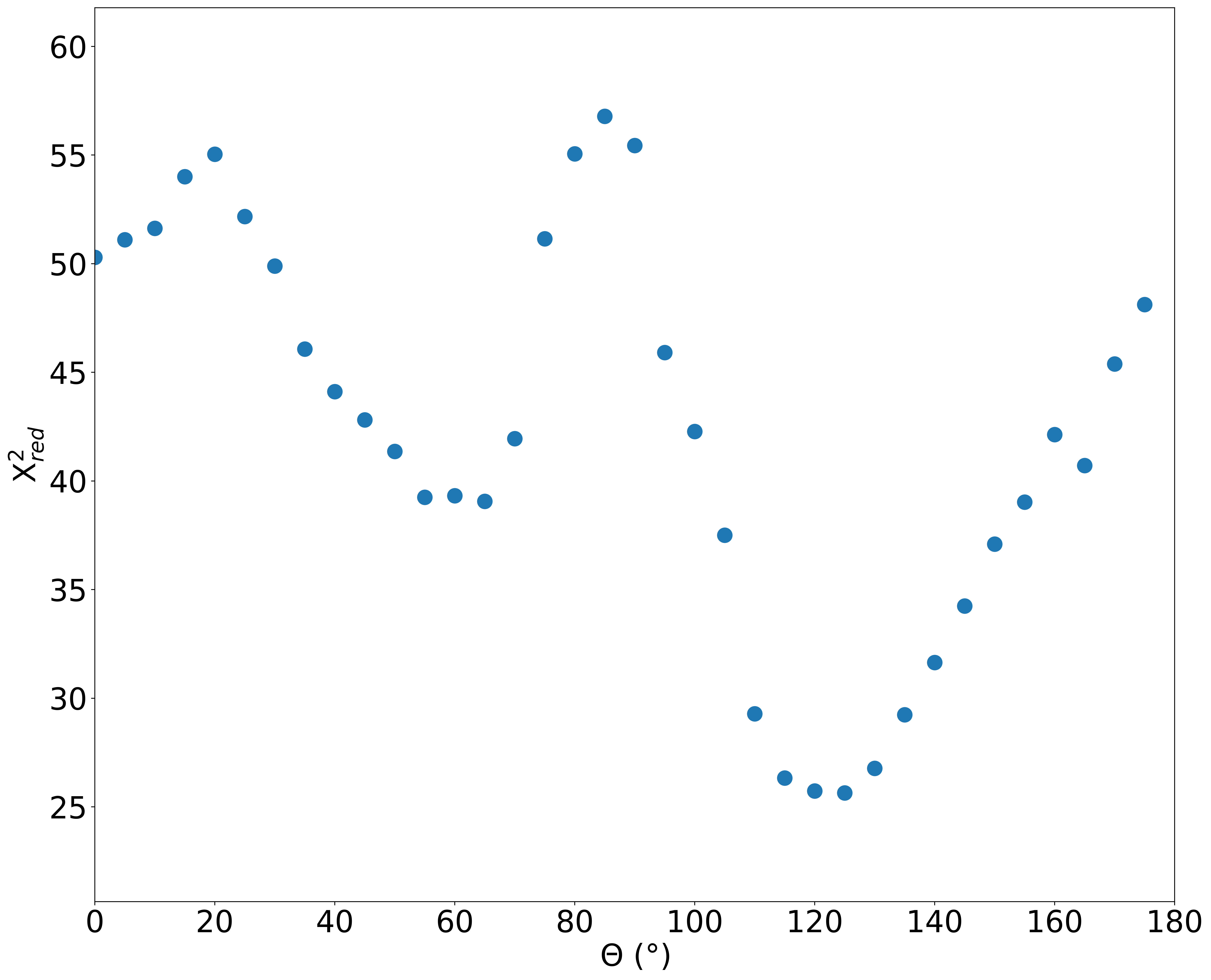}
	\caption{\textit{Left}: The 850 $\mu$m dust emission map is shown with the POL-2 850 $\mu$m polarization vectors ($I >$ 20$\delta_{I}$ and $p >$ 2.5$\delta_{p}$) plotted in black over the best fit dipole morphology shown by the cyan stream-plot (the orientation of the field-lines represent the polarization "E-vectors"). The image is zoomed in to show just the envelope. \textit{Right}: A plot of the reduced $\chi^2_{\rm red}$ value which is found using Eq.~\ref{eq:chi} as it varies with the different field rotations $\Theta$. A minimum can be seen around $\Theta$ = 125\degree\,  (E-vectors) with an uncertainty of $\sim$15\degree\}}
	\label{Fig:chisq}
\end{figure*}

In the fitting, we fix the center of the dipole model at the center of the observed map.  We calculated models on a fixed rectangular grid, over a POS rotation $\Theta$ of 0\degree\, to 180\degree\,, in spacings of 5 degrees. For each case, we find the closest model vector to each observed polarization vector.  We then calculate a goodness of fit parameter \citep[reduced $\chi^2_{\rm red}$;][but see also \citep{andrae2010}]{bevington1992} for that rotation $\Theta$:

\begin{equation}
    \chi^2_{\rm red}=\frac{\sum\limits_{i=1}^n\frac{(\theta_{obs}-\theta_{model})^2}{(1.2\,\delta_\theta)^2}}{N-N_c}
\label{eq:chi}
\end{equation}

\noindent
where $N$ is the number of data points fitted (25 POL-2 vectors passed our S/N threshold), $N_c$ is the number of constraints in the model (here just 1: $\Theta$), $\delta_{\theta}$ is the uncertainty in polarization position angle (see eq. \ref{eq:dtheta}) and $\theta_{obs}$ and $\theta_{model}$ are observed polarization position angle (see eq. \ref{eq:polangle}) and polarization position angle from the model, respectively. The best fit model is overlaid on the observational data in the left-hand panel of Figure \ref{Fig:chisq}.  The resulting reduced $\chi^2_{\rm red}$ plot is shown in the right panel of Figure \ref{Fig:chisq}. We find a clear minimum $\chi^2_{\rm red}$ of $\sim$25.6 when the dipole is rotated to $\Theta_E$=125\degree\, East of North (or -55\degree\, with the normal POL-2 notation). The average measurement uncertainty for the position angles is 11.5\degree\, with a standard deviation of 6.5\degree\,.  The rotation angle is well constrained, with a fitting uncertainty of $\sim$15\degree\, .

\begin{figure*}
	\centering
        \includegraphics[scale=0.52,angle=0]{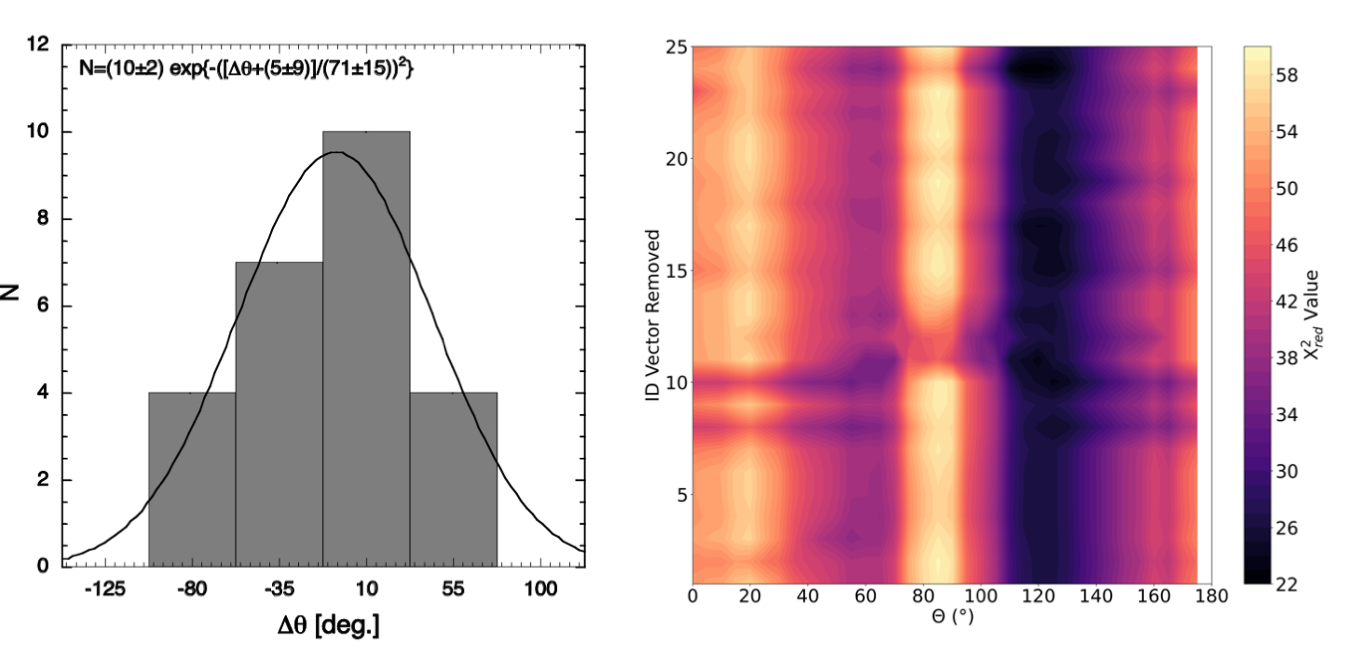}
	\caption{To test the validity of the $\chi^2$ minimization, noting that the reduced $\chi^2$ never approaches zero, we follow \citet{andrae2010}.  (Left) The residuals for the best fit solution yields a peaked distribution.  (Right) Recalculating the $\chi^2$ minimization while eliminating each of the data points, in turn, shows a consistent $\chi^2$ minimum.  The ordinate shows the eliminated data point (arbitrarily numbered). }
	\label{Fig:DPA_hist}
\end{figure*}

The minimum value of reduced $\chi^2_{\rm red}\sim 26$ in our analysis indicates that this is either, over-all not a good fit or, that the model is not a complete description of the data. Given the clear minimum in the $\chi^2_{\rm red}$ we explore its significance with two simple tests \citep[cf][]{andrae2010}.  Figure \ref{Fig:DPA_hist} (left) shows the distribution of residuals between the observed and modeled position angles at the $\chi^2_{\rm red}$ minimum.  While broad, the residuals show a peaked distribution centered on zero (within the uncertainty).  The normalized residuals also show a clear Gaussian distribution, centered at zero, albeit with $\sigma^2>>1$.  Additionally, we performed a jack-knife test \citep{lupton1993,andrae2010}, where we omitted each data point, in turn, and repeated the $\chi^2_{\rm red}$ minimization.  The resulting variation in $\chi^2_{\rm red}(\Theta)$ is shown in the right-hand panel of Figure \ref{Fig:DPA_hist}.  The general shape of the fit-results, and location of the minimum, are consistent for all data-point exclusions.  These results point to a dipole as being a meaningful part of the true, underlying, magnetic field geometry.  More and higher signal-to-noise data, as well as a full radiative transfer model treatment, are required to conclusively address this question.

With the reasonable fit to the 850~$\mu$m data, and since the 154~$\mu$m SOFIA/HAWC+ data at offsets beyond $\sim$30\arcsec\ from the central star show significant deviations from a fully radial polarization pattern, we applied the model also to the 154~$\mu$m SOFIA/HAWC+ observations.  If we combine a `mono-pole' (radial) geometry inside 30\arcsec\ with a dipole model at larger radii for the 154~$\mu$m observations, the vectors in the north-eastern quadrant can be better fit than by a radial geometry alone.  Even so, the over-all reduced $\chi^2$ for the fit with only a mono-pole component is lower than for the combined fit.

\subsection{Comparison of polarization properties in different wavelengths}

\begin{figure*}
	\centering
	\resizebox{8.7cm}{7.1cm}{\includegraphics{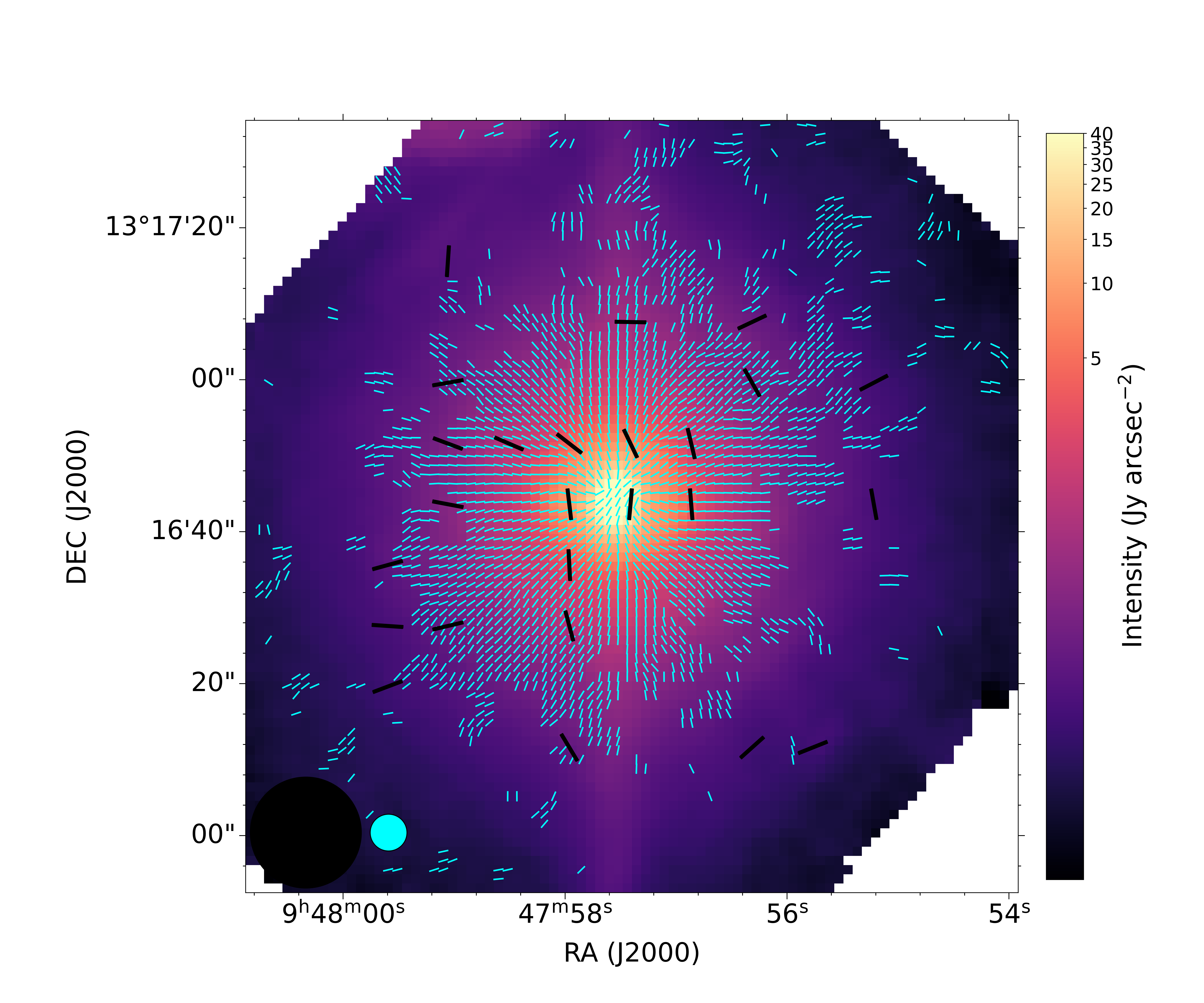}}
	\resizebox{8.7cm}{7.1cm}{\includegraphics{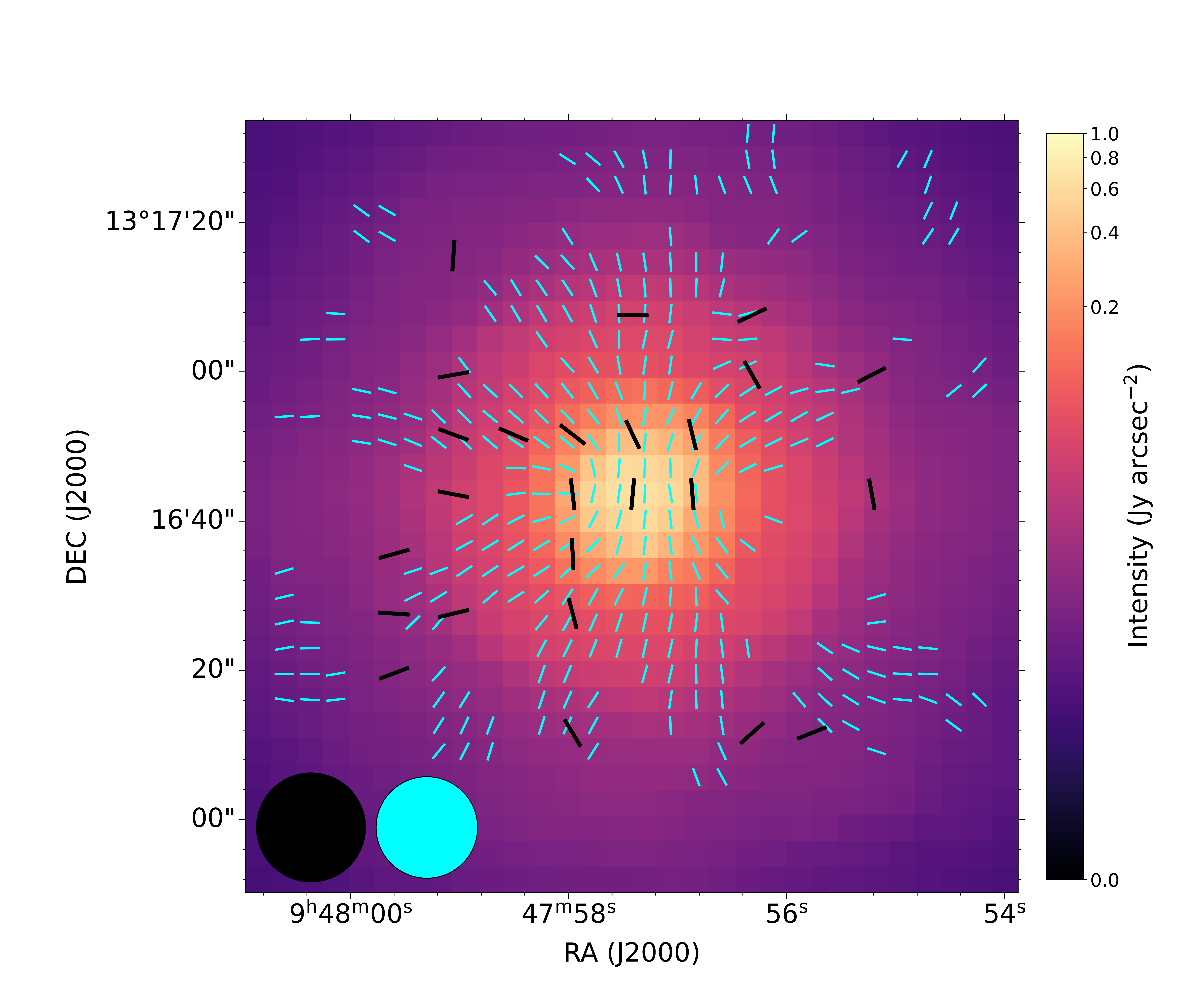}}
	\caption{Left panel shows the SOFIA/HAWC+ 53 $\mu$m polarization vectors in cyan in the envelope of IRC+10\,216 with JCMT/POL-2 850 $\mu$m polarization vectors overlaid in black. Background image is the 53~$\mu$m emission.
	The right-hand panel shows the JCMT/POL-2 850 $\mu$m data, in black, overlaid on SOFIA/HAWC+ 154 $\mu$m emission with polarization vectors (in cyan). Because of the large beam sizes, relative to the offset from the CSE center, the 850 $\mu$m data may be prone to significant beam-averaging and not show the true polarization. In this figure, we show the over-sampled maps of the HAWC+ data to emphasize the differences in polarization geometry. Beam sizes are shown in the lower left corner.} 
	\label{Fig:pol_vectors}
\end{figure*}

The clear differences in the polarization geometry between the 53 and 850 \mum data, makes it is worthwhile to directly compare the polarization at several wavelengths.  Figure \ref{Fig:pol_vectors} compares the polarization geometry for the 53, 154 and 850 \mum data.  The 154 \mum does show a somewhat intermediary geometry, possibly indicating a transition from the radial polarization seen at 53 $\mu$m to the "dipole-like" structure seen at 850 $\mu$m.  We note, however, that while the beam sizes at 154 and 850 $\mu$m are comparable, they are significantly larger than that for 53 $\mu$m, so the comparison is not fully straight-forward.  

\begin{figure}[h]
	\centering
	\resizebox{16cm}{9cm}{\includegraphics{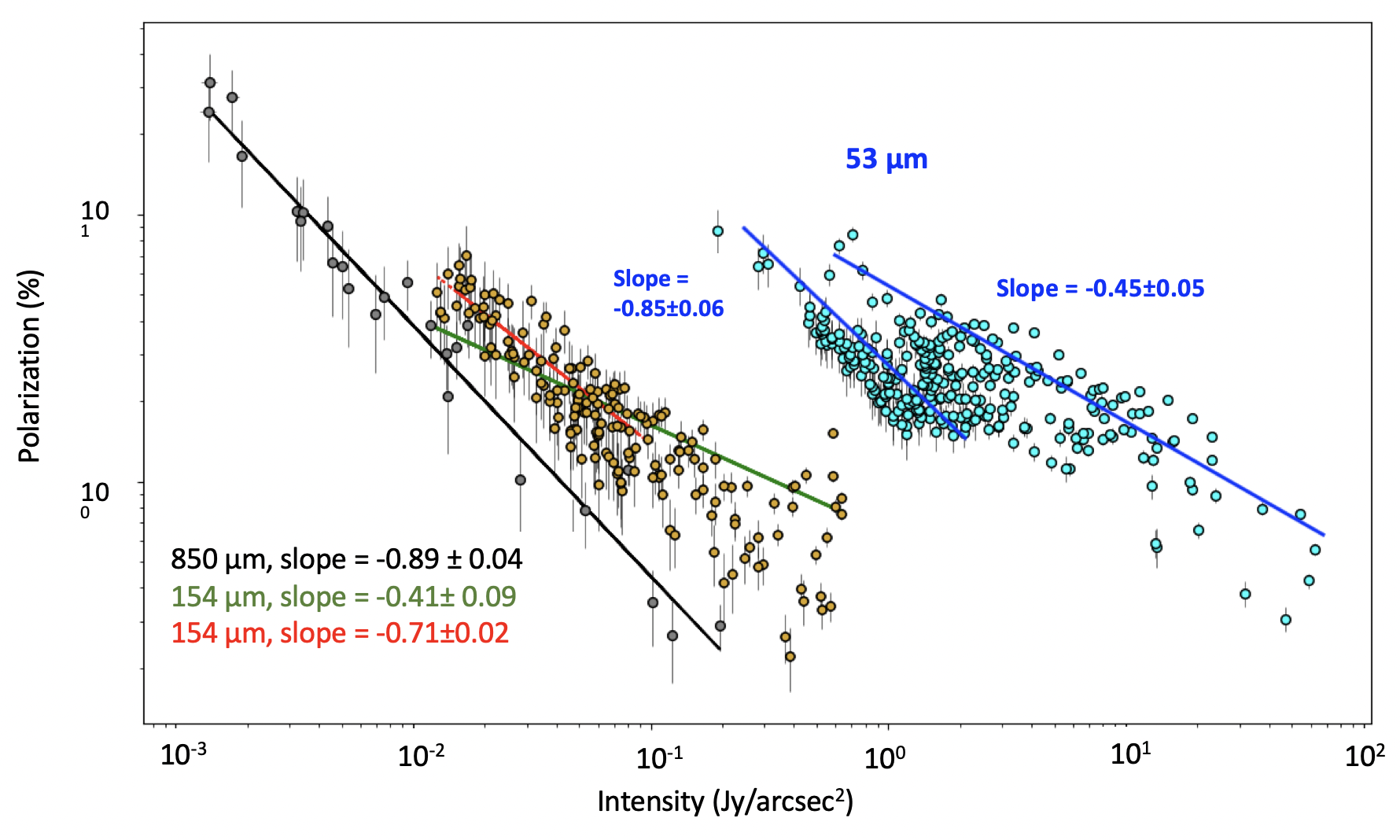}}
	\caption{The polarization fraction for the 53, 154, and 850 $\mu$m polarization is plotted against the Stokes $I$ intensities at each wavelength.  For the 850 $\mu$m data, a single well-defined power-law is apparent.  The 53 $\mu$m is well fit by a two-component distribution with a steeper slope at small fluxes (similar to the 850 $\mu$m one) followed by a shallower slope, offset to somewhat higher values at the intersection.  The 154 $\mu$m falls in between the two, with a hint of an offset, seen in the 53 $\mu$m data, at the high end of the 154 $\mu$m intensity distribution.  If the 154 $\mu$m data are fit to only intensities less than 0.1 Jy/arcsec$^2$ an exponent consistent with that for the steep part of the 53 $\mu$m and the 850 $\mu$m data is found.  See text for details.  All data have been debiased. }
	\label{Fig:polfrac}
\end{figure}

Figure \ref{Fig:polfrac} shows the polarization fractions at 53, 154 and 850 $\mu$m, as functions of the total intensities. A similar drop from high polarization and low intensities at larger radii to low polarization and high intensities near the star can clearly be seen for all wavelengths. Fitting single component power-laws, the drop is steeper for 850 $\mu$m, with a slope (exponent of a power-law) of -0.89$\pm$0.04.  Shallower, single-slope relations are found for 154 $\mu$m (slope= -0.41$\pm$0.09) and for 53 $\mu$m (slope= -0.31).  As is clear from Figure \ref{Fig:polfrac}, a single component power-law may, however, not be adequate, especially for the 53 $\mu$m data.  A two power-law fit to the 53 $\mu$m data yields an initial slope consistent with that for the 850 $\mu$m data (slope= -0.85$\pm$0.06), with a slope of -0.45$\pm$0.05 at higher intensities, slightly offset to higher polarization fractions.  The 154 $\mu$m polarization shows an indication of a similar offset/second-slope at the highest intensities in that band (I$\geq$0.2 Jy/arcsec$^2$).  This potential second distribution is most strongly indicated by the fact that if the 154 $\mu$m data were fitted to only intensities less than $\sim$0.1 Jy/arcsec$^2$ a significantly steeper power-law would have resulted (slope= -0.71$\pm$0.02).  
The "jump" in the 53 $\mu$m polarization occurs at $\sim$3 Jy/arcsec$^2$.  As Figure \ref{Fig:I_vs_r} shows, the two intensities occur at similar radial distance from the central star of $r\approx 12.5\arcsec$.  We note that at this radius, the Stokes $I$ intensity for the 850 $\mu$m emission is $\sim$6.5 mJy/arcsec$^2$ and therefore well beyond the current polarization sensitivity.  Hence, the polarization fractions in each band vary in similar manners with a steeper slope at large distances in the CSE and shallower slopes inside about 12.5\arcsec.  This may indicate a change in alignment mechanism within the shell, or a reduction in the alignment efficiency (including disalignment effects) closer to the star.  While the higher gas density and dust temperature closer to the central star would both, nominally, imply larger collisional and radiative disalignment \citep[cf][]{draine1998}, we would not expect them, by themselves, to be able to explain the discrete change in slope of the polarization.  Both the radiation field and gas density in CSEs are expected to be continuous and vary smoothly with radius.  Although significant, small scale, structure is known to exist in the CSE \citep[eg.][]{mauron1999,leao2006,ladjal2010}, the general r$^{-2}$ radial density dependence will still dominate the radial density profile.  Hence, the observed discrete changes in polarization efficiency, likely require a more qualitative change in the grain alignment, such as a different mechanism.

\subsection{Line Polarization}\label{Sec:line_pol}

As noted above, Goldreich-Kylafis polarization \citep{goldreich1981,goldreich1982} has been observed in several sub-mm wave lines in the envelope of IRC+10\,216 \citep{girart2012}.  They mapped the linear polarization from the lines of CO (J=3–2) at $\nu$=345.796 GHz, SiS (J=19–18) at $\nu$=344.779 GHz, and CS 7–6 at $\nu$=342.883 GHz.  These lines are all located at wavelengths with transmissions above 95\% of the peak transmission of the SCUBA-2 850 $\mu$m filter \citep{cookson2018}.  The SMA primary beam (6m antennae) at 850 $\mu$m is about 36\arcsec and \citet{girart2012} achieved a synthesized beam of 2.6\arcsec$\times$1.6\arcsec.  Also, as shown by \citet{girart2006}, the SMA filters out emission that arises from structures larger than about 10\arcsec, so the two data sets are not trivial to compare in a quantitative way.  However, if we use the CO data from \citet{girart2012} and the conversion factors given in \citet{drabek2012}, we can estimate the possible Goldreich-Kylafis contamination in our observations.  Using the \citet{girart2012} measurement of the Stokes $Q$ parameter, we find a peak line brightness temperature of 2 K and a width of 20 km/s. If the JCMT conversion factor is valid also for the Stokes $Q$ parameter, we find a polarized intensity on the order of 30 mJy/beam which would contribute a significant signal in our polarized intensity, assuming the spatial filtering does not strongly affect the comparison.  

\begin{figure}
\centering
\resizebox{18cm}{14cm}{\includegraphics{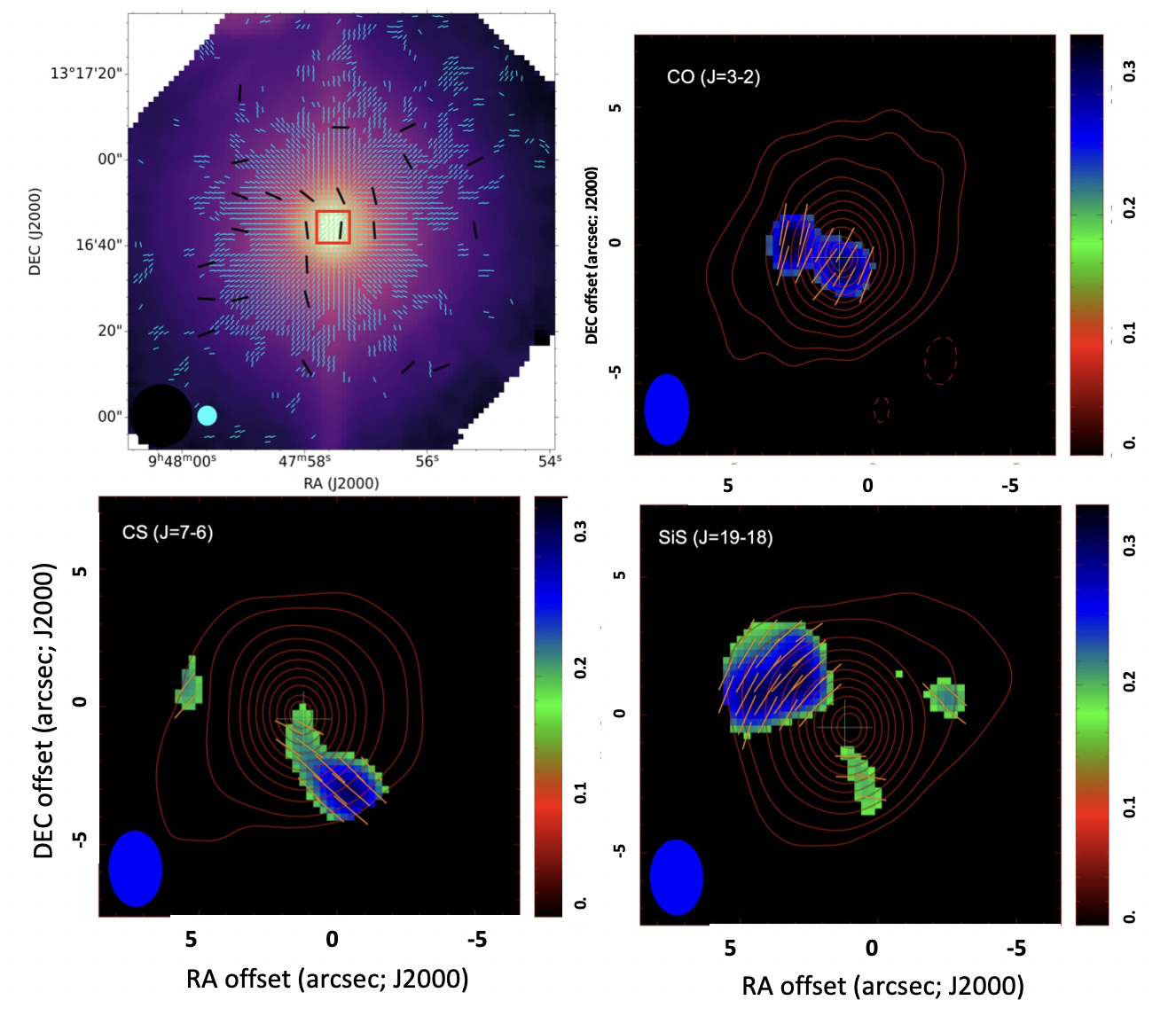}}
\caption{Comparison between the (upper left) JCMT/POL-2 850 $\mu$m (black vectors) and SOFIA/HAWC+ 53 $\mu$m (cyan vectors) polarization with the line polarization observation by \citet{girart2012} (reproduced with permission).  The red square in the upper-left map shows the approximate extent of the three sub-figures from \citet{girart2012}.  In the three panels from \citet{girart2012} the linearly polarized intensity is shown as color plots,  while the contours represent the Stokes $I$ emission (see \citet{girart2012} for further details). While the JCMT/POL-2 data at this scale is prone to some beam averaging, the polarization geometries between the continuum and line polarization are not consistent with a common origin.  Note that the caption of Figure 2 in \citet{girart2012} mislabels the SiS and CS images.}\label{Fig:JCMT_vs_SMA}
\end{figure}

However, if we compare the polarization geometry between our observations (SOFIA/HAWC+ and JCMT/POL-2) with those of \citet{girart2012} the picture becomes more complicated.  As Figure \ref{Fig:JCMT_vs_SMA} shows, the latter only detect emission in the very inner part of the CSE and our field of view, where we might expect beam averaging effects in  the JCMT/POL-2 data.  Nonetheless, none of the CO, SiS or CS line emission data show polarization geometries that are consistent with those seen in the continuum, either at 53 or 850 $\mu$m.  In addition, the position angle of the CS (J=7-6) emission (PA$\sim48^\circ$) is close to perpendicular to those for CO (J=3-2) (PA$\sim155-170^\circ$) and SiS (J=19-18) (PA$\sim139^\circ$), and the different line polarization components would therefore offset each other within the broad SCUBA-2 band-pass filter (and larger beam).

\subsection{Grain Charging}

Charged grains can interact with a surrounding magnetic field, even for diamagnetic grains.  As we discuss below (Sec. \ref{Sec:E-RAT}), a recently formulated grain alignment mechanism \citep["E-RAT"][]{lazarian2020}, predicts that grains not susceptible to paramagnetic interactions can achieve efficient alignment with the magnetic field, if they are charged and posses a relative motion to the field.

\begin{figure*}[h]
	\centering
	\resizebox{17cm}{7.5cm}{\includegraphics{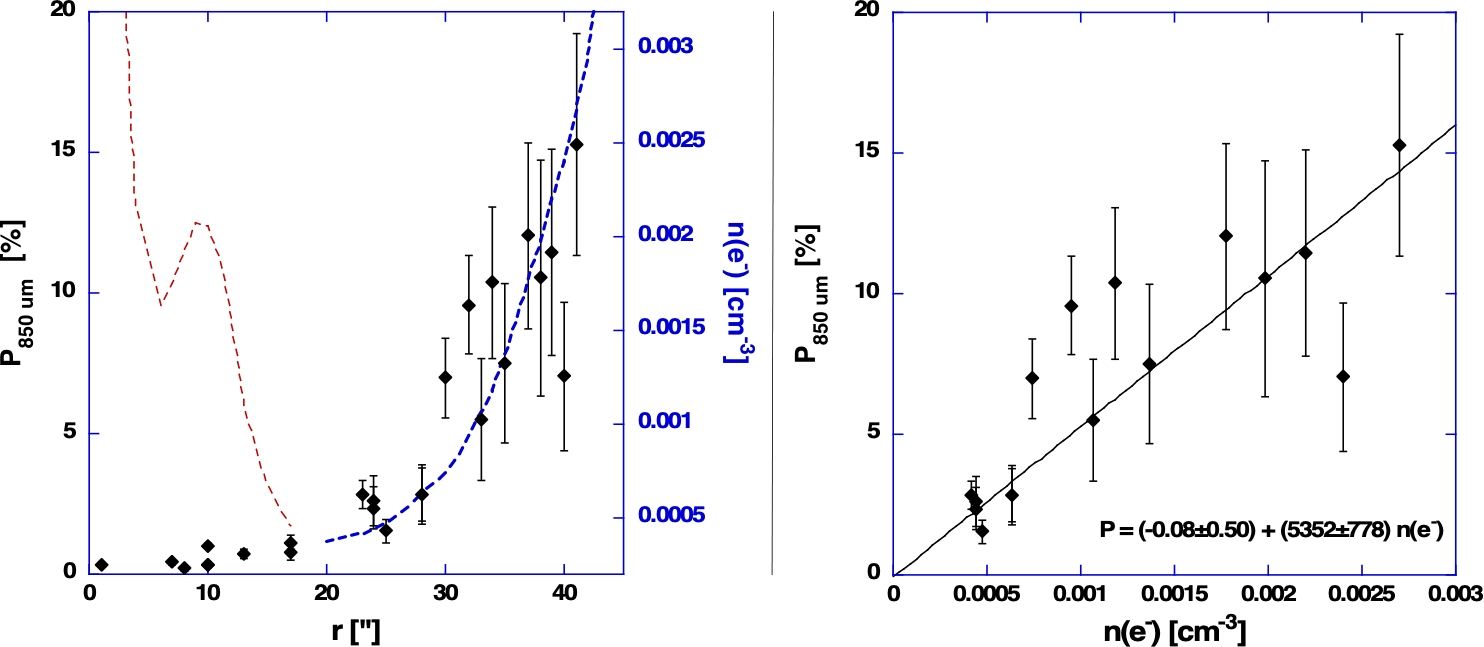}}
	\caption{(Left) The amount of 850 $\mu$m polarization is well correlated with the electron density by photo-ionization due to the external diffuse radiation (UV) field (blue dashed line) calculated by \citet{li2014} in the IRC+10\,216 envelope (the scaling of the two vertical axes have been arbitrarily set to show the correlation). The red dashed curve represents modeled electron density generated by cosmic rays in the denser inner shell. As shown by \citet{ivlev2015} grain-charging by cosmic rays in the inner envelope (similar to their region "$\mathcal{I}$") is inefficient. (Right) A direct comparison between polarization and electron density for distances beyond 20\arcsec\ (where the latter is dominated by photo-ionization in the \citet{li2014} models) yields a formally significant correlation.}
	\label{Fig:p_vs_e}
\end{figure*}

For dense AGB star CSEs the photo-ionization of the gas, and charging of the dust, is expected to be strongest in the outer envelope and originate from the UV light in the diffuse interstellar field (because of the cool effective temperature of the central star very little UV flux is expected from its photosphere).  Since we do not have any direct observational tracer of the grain charging, we used the gas electron density calculated in the detailed chemical modeling of IRC+10\,216 by \citet{cordiner2009} and \citet{li2014} as a proxy for the grain charging as a function of radius. Figure \ref{Fig:p_vs_e} shows the comparison of polarization fraction to the electron density.

For interstellar conditions, where grain-electron collisions are an important aspect of the grain charge balance \citep{weingartner2001}, we would expect the grain charge to be proportional to 1/n(e).  However, the conditions in the CSE (low fractional ionization, fully molecular hydrogen, high flow-speed and steep density gradient, etc.) may mean that the collisional effects are negligible, and the grain charge will be determined by photo-electric emission balanced by the electrostatic potential of the grain, and hence be dependent on the opacity to the interstellar radiation field, but not involve the 1/n(e) factor, as implied by the semi-empirical correlation seen in Figure \ref{Fig:p_vs_e}.  Detailed modeling, applicable to the conditions of the CSE (beyond the scope of this paper), will be needed to fully address this issue.

While for small grains the electric moment can originate as part of grain structure \citep{draine1998}, for the large grains indicated in the IRC+10\,216 CSE \citep[e.g.][]{groenewegen1997} grain charging is likely required (A. Lazarian, 2022, private communications). Since the electric dipole moment of a grain depends on the total charge on the grain \citep[cf.][]{weingartner2006,lazarian2020}, we use the space density of electrons as our proxy for grain charging and electric alignment.  The blue dashed line in the outer envelope corresponds to electrons released by photo-ionization, while the red dashed curve corresponds to electrons due to cosmic ray ionization.  \citet{ivlev2015} have argued that grain-charging by cosmic rays in the inner envelope, which is similar in gas parameters to their region "$\mathcal{I}$", should be inefficient.

\section{Results}\label{sec:res}

Our FIR, especially the 53 $\mu$m, and optical polarization data \citep[][Andersson et al., 2023b, in preparation]{bga2022a} indicate grain alignment with the long axis in the radial direction away from the central star, supporting a second-order radiative grain alignment mechanism \citep{hoang2009a}.  We stress that, as we argued in the former paper \citep{bga2022a}, the radial dependence of the FIR polarization does not support a mechanical (gas-dust drift) alignment mechanism \citep[e.g.][]{gold1952} as the primary driver.  The JCMT/POL-2 850 $\mu$m data, presented here, do, however, not show this simple radial pattern, even in the inner part of the CSE.  

The analysis utilizing the radial coordinate polarization (Sec. \ref{sec:rad_pol}) shows a distinct difference between the FIR (SOFIA/HAWC+) and sub-mm wave (SCUBA-2/POL-2) data.  Again, for the FIR, the $\cos{\theta_r} >0.95$ (for all 53 $\mu$m data and inside $r \approx$ 32\arcsec\ for the 154  $\mu$m data) indicate a clear radial polarization pattern (Figure~\ref{fig:Radial}, upper panels).  While of lower signal-to-noise, and at somewhat different scales, the plots indicate neither a radial nor a tangential pattern in the sub-mm wave data in Figure~\ref{fig:Radial}, lower panel.


Using the Zeeman measurement results of \citet{duthu2017}, we find a periodic variation in the line-of-sight magnetic field around the star, with the pole of the possible underlying dipole field oriented at around 90-125\degree\,, albeit with relatively large uncertainties.  We therefore assume a multipole expansion of a potential magnetic field and apply a simple dipole model fit to the 850 $\mu$m data.

As Figure~\ref{Fig:chisq} shows, there is a clear $\chi^2$ minimum found for a rotation angle of $\Theta_E$=125\degree\.  The modeling is done for E-vectors.  The relationship to a possible magnetic dipole field axis is ambiguous, depending on the alignment mechanism, even if involving magnetic fields (see Sec. \ref{Sec:E-RAT}).  The polarization symmetry axis is close to that of the Zeeman results and of the torus seen by \citet{jeffers2014}. 

While our simple model uses only a \textit{projected} dipole term for a possible magnetic field and does not fully account for all measured orientations of the polarization field, it is clear that the FIR and sub-mm wave data do not show the same (dominant) polarization patterns.

Comparing to the line polarimetry by \citet{girart2012} shows that, while significant, Goldreich-Kylafis polarization of spectral lines within the SCUBA-2 pass-band is unlikely to explain the observed 850 $\mu$m continuum polarization (Sec. \ref{Sec:line_pol}).

As shown by Figure \ref{Fig:polfrac}, the slope (power-law exponent) of the fractional polarization as a function of total intensity is nominally steeper for the 850 $\mu$m polarization, than for the 53 \mum or 154 $\mu$m ones.  Allowing a two-power-law fit to the 53 (and 154) $\mu$m data, however, shows that at small respective intensities all three share a common slope of approximately -0.85.  The second power-law parameters are only measurable for the 53 $\mu$m data, yielding a slope of -0.45, with an offset to higher polarization fractions for a given intensity, with the transition between the two at $\sim$3 Jy/arcsec$^2$.  The 154 $\mu$m polarization shows an indication of a similar change at $I \approx$ 0.2 Jy/arcsec$^2$.  These intensities both occur at a radial offset from the star of $\sim$12.5\arcsec\ (Figure \ref{Fig:I_vs_r}).  While this is comparable to the beam size at 154 $\mu$m, it is significantly bigger than the 53 $\mu$m beam. 

 Figure \ref{Fig:p_vs_e} shows a correlation between grain charging and alignment in the outer envelope of IRC+10\,216 beyond an offset radius of $\sim$20\arcsec.

\section{Discussion} \label{sec:Disc}

As shown by our polarization maps and the azimuthally averaged radial polarization plots, the 850 $\mu$m polarization is dominated by neither a radial nor an azimuthal pattern.  The small number, and asymmetric distribution, of the JCMT polarization detections make the difference between 53 $\mu$m and 850 $\mu$m data difficult to interpret conclusively.  However, the non-radial polarization in the latter data set does require a different polarization mechanism than we have proposed for the FIR data.

For the inner part of the CSE, the large beam size (14\arcsec), relative to the offset from the center of the CSE, means that the JCMT data is prone to, possibly severe, beam-averaging effects.  However, as can be seen from Figure \ref{Fig:pol_vectors} the beam-averaged polarization even inside $\sim$20\arcsec of the central star is often azimuthal (or, at least, non-radial) in orientation.  By symmetry, a beam-averaged radial pattern would remain radial, and therefore it is unlikely that the beam size can explain this dichotomy between the FIR and sub-mm wave polarization for dust at a (3D) radial distance from the central star comparable to the projected distance of these measurements.

If the more efficient 850 $\mu$m polarization is due to the dust, then it indicates that some of the grains have acquired efficient internal dissipation, to be aligned by some version of the RAT mechanism, or are aligned by the RAT-like mechanical alignment (MET) mechanism discussed by \citet{lazarian2021}, where gas particles reflect off of a helical grain surface and therefore align the grain's angular momentum vector with the gas-dust drift velocity.  MET alignment would therefore, as would first-order k-RAT, produce azimuthal polarization in the AGB star envelope.  However, given that only a small number of points show close to azimuthal polarization, neither first order k-RAT nor MET dominated alignment seem likely. As noted by \citet{hoang2009a} weak "direct RAT alignment" can be achieved along the $B$-field also without internal alignment which for low-J attractor points can be with the long axis along the magnetic field.

Some level of internal alignment may be achieved from inelastic dissipation in porous grains \citep[][Tram, 2022, private communications]{purcell1979}. This seems somewhat unlikely in the IRC+10\,216 CSE, given the need to preserve large grains \citep{groenewegen1997} under strong radiative disruptive driving \citep{hoang2019} which, for porous grains with small tensile strength, would result in the large aligned grains being destroyed.  We note that the grain disruption is limited to grains at high-J attractor points \citep{lazarian2021}, and a significant fraction of the dust may therefore not be affected by the disruption.

We are therefore left with magnetically aligned grains, possibly due to mineralogy changes or driven by grain charging and induced electric fields \citep{lazarian2020} or some as yet unidentified grain alignment mechanism.  Contamination by polarized line emission (Cernicharo 2021, private communications) also cannot be fully discounted, at this stage, as discussed above. 

Given the differences in polarization geometry between FIR and sub-mm wave data, we used the Zeeman observations from \citet{duthu2017} to search for periodic variability in the line-of-sight magnetic field around the central star.  While there are only five data points (plus an upper limit) we find a good fit to a cosine function of the azimuthal angle, $\Theta$ around the star with the extremum located at $\sim$90-125$^{\circ}$. We interpret this as indicating a dipole pattern (being the second order component of a spherical harmonics expansion). A dipole with its symmetry axis in the plane of the sky will not show any net line-of-sight polarization, since in the equatorial plane the $B$-field is in the plane of the sky, while towards the poles there will be equal contributions of positive and negative fields along the line of sight (assuming optically thin emission).  However, if the symmetry axis is tilted out of the plane of the sky, a net line-of-sight magnetic flux will result \citep[as seen by e.g.][]{vlemmings2005}.   

As our modeling (Figure \ref{Fig:chisq}) shows, a simple projected dipole model rotated by 125$^\circ$ east of north (E-vectors) yields a distinct nominal $\chi^2$ minimum in the fit of the 850 $\mu$m data, with a symmetry axis consistent with that implied by the Zeeman data.  This provides some support for B-RAT alignment of the grains, or some other magnetically defined alignment mechanism giving rise to the 850 $\mu$m polarization.  As noted above, the large absolute minimum value of the reduced $\chi^2$, means that a projected dipole is not, by itself, a complete or accurate model of the data.  However, the localization of the symmetry axis seems to indicate that it likely forms part of the true polarization geometry.

It is noteworthy that \citet{jeffers2014} have modeled the scattered light in the inner $\sim$3\arcsec of the IRC+10\,216 CSE as due to a torus centered on the star, with a symmetry axis of $\sim$75$^\circ$ east of north.  In our optical data, the only stars whose polarization diverges significantly from being azimuthal with respect to the central star IRC+10\ 216 intercept this symmetry axis.   If this represents the angular momentum/rotation axis of the CSE, it would imply a close alignment with the indicated magnetic axis.

\subsection{Polarized Line Emission Contamination}\label{sec:disc_linecont}

While the nominal polarized line flux in the G-K polarization is similar to the SCUBA-2 polarized flux, the geometry of the two types of polarization do not seem consistent.  Because the spatial sampling of the SMA and the JCMT are significantly different, we cannot conclusively rule out a measurable line polarization contamination in our data, but a dominant contribution from G-K polarization seems unlikely, because of the geometry.  Simultaneous line and continuum polarization of IRC+10\,216 with ALMA are planned for Cycle 9 to better address this issue.

\subsection{Magnetic Dust Polarization Mechanisms}\label{sec:disc_polmech}

For dust polarization to be tracing a magnetic field in the CSE, the grains dominating the 850 $\mu$m polarization must have acquired at least the equivalent of paramagnetic characteristics. This might be caused by variations in the grain chemistry with radius from the star, or with grain size (or some combination of these).

First, when considering the grain chemistry, mineralogy of AGB envelopes naturally show a radial variation due to the evolution on the AGB. As noted above, when a star first enters the AGB it is oxygen rich.  As the thermal-pulse driven dredge-up of newly synthesized material is mixed up to the stellar surface and ejected into the CSE it eventually becomes carbon rich.  This may leave silicate grains in the outer part of the CSE, with carbonaceous dust in the inner part of the envelopes, as reported for the planetary nebula BD+30$^\circ$3639 \citep{guzman-ramirez2015}.  The JCMT/POL-2 850 $\mu$m observations may therefore be tracing this outer, colder, fossil silicate dust seeing only dust at large distances from the star, even when projected close to the central star.  However, because the emission is optically thin, the dominant contribution also to the 850 $\mu$m Stokes $I$ emission originates in the denser gas at relatively small radii away from the central star.  Given the expansion velocity of 14 km/s and a distance of 123 pc, the presence of such fossil silicates would also require that the transition from oxygen to carbon rich chemistry would have been very recent, or that the mixing of the phases be very efficient \citep[cf][]{leao2006,sahai2010}.

The possibility for explaining efficient alignment due to grain size, for the grains dominating at 850 $\mu$m, parallels the proposed explanation of the shape of the optical polarization - "Serkowski" - curve under Davis-Greenstein (henceforth "DG") alignment by \citet{mathis1986}, using super-paramagnetic grain inclusions.  The empirical Serkowski curve for diffuse ISM lines of sight requires that only grains with radii larger than $\sim$0.045$\mu$m are aligned \citep{kim1995}.  RAT alignment explains this as due to the lack of radiation below the Lyman limit in neutral gas (the condition for efficient RAT alignment being $\lambda <$ 2\textit{a}, where \textit{a} is the effective grain radius).  Because DG alignment becomes more efficient for smaller grains \citep[e.g.][]{draine2011}, a mechanism was needed to deactivate it for grains smaller than this size. Because DG alignment further requires significantly higher magnetic susceptibility in the grains than for ordinary silicates \citep{jones1967}, Mathis proposed that the lack of small grain alignment was due to the fact that grains get aligned only if they contain at least one "super-paramagnetic inclusion" (of e.g. metallic iron, FeO, FeS, etc., \citep[cf][]{goodman1995}).  While mid-infrared observations show that the dust in the IRC+10\,216 CSE is fully carbonaceous (95\% amorphous carbon and 5\% SiC; \citet{ivezic1996a}), iron is found to be depleted in the CSE gas \citep{mauron2010}.  It is therefore possible that the difference between FIR and sub-mm wave polarization originates in differences in the alignment properties of smaller (and hotter) grains seen in the MIR/FIR, and larger (cooler) ones seen in the sub-mm, where the latter have acquired inclusions of e.g. ferromagnetic metallic iron or iron carbide \citep[Fe$_3$C;][]{wang2017}. Since this is only a change in the alignment efficiency, it would not affect the Stokes $I$ profile.

The fractional polarization, as a function of Stokes $I$ intensity, shows a bimodal dependence in the 53 $\mu$m data, with a hint of a similar behavior at 154 $\mu$m, indicating a change in alignment behavior at about 12.5\arcsec\ from the star.  Since the slope at higher intensities (closer to the star) is shallower and the absolute level is higher for this distribution than the steeper slope extending further out, this is unlikely to be due to radiative dust disruption \citep{hoang2019, hoang2021}.  Quantitative modeling will be needed to explore the origin of this bimodality, but does support a change in grain alignment physics with radius in the CSE.

\subsection{Electrically Induced Alignment}\label{Sec:E-RAT}

Because of the cool photosphere of CW Leo\footnote{For clarity, we refer to the central star by its name when specifically addressing stellar parameters, and "IRC+10\ 216" for the extended object, including the star} and the high column density of the CSE, the inner part of the envelope is expected to be close to neutral \citep[modulo cosmic ray ionization;][]{cordiner2009}, as supported by the strong absorption in \ion{K}{1} (ionization potential of 4.3 eV) seen in "Star 6" of \citet{mauron2010}, even at an offset of 37\arcsec\ from 
the center of IRC+10\,216.  

There is growing evidence for some kind of UV source at or close to the center of the CSE.  Ionized carbon (through the 158 $\mu$m line of [\ion{C}{2}]) has been observed with Herschel/HIFI toward the center of the CSE \citep{reach2022}, as has high-J line of HC$_3$N with ALMA \citep{siebert2022}, which both require UV photons for their production.   Based on imaging observations by \citet{kim2015,kim2021} and variability studies by e.g. \citet{cernicharo2015,decin2015,guelin2018}, \citet{siebert2022} added a central UV source - most likely a "Solar-like companion" - to their chemical modeling of the HC$_3$N abundance.  The inner density peak of HC$_3$N resulting from this assumption is, however, located within a small enough radius ($\lesssim$ 5\arcsec\ ) that our observations would be significantly beam averaged in this region.

For the outer parts of the CSE, UV photons from the diffuse interstellar field can penetrate the medium and ionize the gas, as well as charge the grains.  While the work function for interstellar grains is not well known, the value of $W \approx$4.4 eV, appropriate for graphite \citep{weingartner2001}, is lower than the ionization potential of both hydrogen and carbon.  Since both the grain charging and ionization fraction in the gas are due to the external UV field, we will therefore, here, use the gas electron density as a proxy for grain charge. As discussed above, the validity of this assumption depends on grain-electron collisions to be negligible.  Figure  \ref{Fig:p_vs_e} shows a clear correlation between the polarization fraction and electron density at offsets $\gtrsim 20\arcsec\ $from the center of
IRC+10\,216.

As discussed by \citet{ivlev2015} and \citet{ibanez-mejia2019}, while the radiative grain charging at low extinctions for carbon grains is efficient, due to their low work function, the cosmic ray grain charging (mostly through secondary H$_2$ UV emission in the Lyman and Werner bands) is less so, by at least an order of magnitude (see Figure 3 of \citet{ibanez-mejia2019}).  This, possibly together with the beam averaging at the very center, likely explains the lack of correlation between $n$(e$^-$) and $P$ inside $r \approx 20\arcsec$.

For emission polarimetry the polarization fraction $P$ inherently has the appropriate line of sight averaging needed to address alignment effects (as noted by \citet{hildebrand2000} for \textbf{ext}inction and \textbf{em}ission polarization, generally: $P_{\mathrm{em}} = - P_{\mathrm{ext}}/\tau$, where the minus sign denotes the fact that the emission and extinction polarization position angles are orthogonal.)  No significant correlation is seen between $n$(e$^-$) and the polarized intensity; $IP$.  Because of the non-uniform density structure (and spherical symmetry) of the CSE, our comparison is, of course, still only approximate, but points to a likely correlation between grain alignment and grain charging, in the outer envelope.

As discussed above, \citet{lazarian2020} predicts that charged carbon grains with a net velocity perpendicular to the magnetic field should align with the magnetic and electrical fields (since the direction of the induced E-field is given by the direction of the $B$-field).  If the precession rate around the E-field ($\Omega_E$) is faster than around the radiation field direction ($\Omega_k$) and the B-field ($\Omega_B$), then theory predicts that the grain should align with their long axis parallel with the B-field and thence the observed emission polarization should be parallel to the $B$-field.  Since the polarization field (E-vector) symmetry, from our modeling, agrees with that from Zeeman data, our results indicate that this is the case ($\Omega_E > \Omega_B$). If confirmed, these results can therefore provide constraints on the charging, drift velocity, and geometry of the grains.

\vspace{8pt}

To understand the dichotomy between FIR and optical polarization on the one hand and sub-mm wave continuum polarization on the other in the envelope of IRC+10\,216, sub-mm observations at higher spatial resolution and better signal-to-noise are needed. Also, line and continuum polarimetry in the sub-mm wave range acquired with the same instrument would be highly valuable to understand the relative contributions of dust and Goldreich-Kylafis polarization.  The 7-m array of ALMA (and eventually the Morita array) can reach 4\arcsec\ resolution for polarimetry at 870$\mu$m - closely matching that of the SOFIA/HAWC+ 53 $\mu$m observations - and the high signal-to-noise needed to sensitively probe the cold dust in the inner [projected] part of the CSE.

The 3.4 $\mu$m aliphatic CH line is likely too inherently weak to be detectable in polarization in CSEs (\citet{chiar2000} reports line opacities of $\tau \sim$0.2 towards the Galactic center for visual extinctions exceeding 30 magnitudes).  However, the solid-state SiC line at 11.3$\mu$m has been reliably detected in several carbon rich CSEs \citep[e.g.][]{sloan1998}.  Unfortunately, very little polarization data are available to specifically probe the alignment of such grains.  The polarization data of \citet{smith2000} may show a hint of polarization in this line for the carbon-rich post-AGB star AFGL 618 (showing a shape consistent with the flux spectra of carbon stars from \citet{sloan1998}.  While the amplitude of this polarization line in AFGL 618 is only marginal with respect to its uncertainties, the possible detection of SiC polarization suggests that a systematic study of polarization in the line would be very valuable to further probe carbonaceous grain alignment.

\section{Conclusions} \label{sec:concl}

We present 850 $\mu$m JCMT/SCUBA-2/POL-2 polarimetry of the circumstellar envelope of the carbon rich AGB star IRC+10\,216.  In contrast to FIR 53 $\mu$m and optical polarimetry, which show an ordered radial (azimuthal) polarization pattern in the FIR (optical), implying dust grains aligned with their long axis in the radial direction away from the star, the sub-mm wave polarization is more complex.  Our data are not sufficiently extensive to allow a complete modeling of the 850 $\mu$m polarization data.  However, based on an analysis of the Zeeman effect measurements in the CN (J=1-0) line by \citet{duthu2017} and an assumption of the validity of a multipole decomposition of the magnetic field, we fitted a rotated, projected, dipole to the 850 $\mu$m polarization data.  Under this model assumption, a distinct preference for a symmetry axis is found, consistent with both the symmetry axis of the Zeeman results and the dust torus reported by \citet{jeffers2014}.  These results imply that, if the polarization is due to aligned dust grains, the dust traced by sub-mm wave observations is aligned with a magnetic field.  The optical and FIR polarization, in contrast, indicate magnetically inactive dust aligned via a second-order pure radiative mechanism.  To reconcile this dichotomy we consider several possible explanations:  
\begin{itemize}
    \item The 850 $\mu$m continuum polarization may be dominated by Goldreich-Kylafis line polarization e.g. in the lines of  CO (J=3-2), SiS (J=19-18) and CS (J=7-6), contained within the SCUBA-2 pass band.
    \item The cold dust probed by our 850 $\mu$m observations may have a different dust mineralogy than the dust probed by the optical and FIR observations.  We hypothesize that this difference can arise from one of two processes: 
    \begin{itemize}
        \item  The sub-mm wave data may be probing the fossil silicate dust from the oxygen-rich phase of the AGB star, located in the outer parts of the CSE, or 
        \item The largest carbon grains may have acquired magnetically active inclusions from the depleted metals in the CSE (e.g. metallic iron or Fe$_3$C).
    \end{itemize}
    \item A new alignment mechanism \citep{lazarian2020} applicable to charged grains with small magnetic susceptibilities (i.e. carbon grains) may explain our results, as the polarization fraction in the 850 $\mu$m data are well correlated with the calculated electron densities \citep{li2014} in the envelope.
\end{itemize}  

Based on an Occam's razor argument, the latter possibility is particularly attractive, as the FIR and sub-mm wave polarization difference then is only a matter of differences in the grain charging state.

\vspace{8pt}

If the magnetic alignment of the grains seen at sub-mm wavelengths, and the dipole geometry of the resulting polarization, can be confirmed, the large scale structure of the magnetic field around IRC+10\,216 will be known.  The connection to the dust structures seen at small scales \citep{jeffers2014} are particularly intriguing.

Further continuum and line data, at higher resolution and sensitivity, preferably also at intermediate wavelengths are needed to confirm and further explore these results.
Our results indicate that pure carbon grains are likely not aligned in the ISM, and thereby provide observational constraints on the properties of composite grains \citep[e.g.][]{draine2021}.

\begin{acknowledgements}
B-G A., A.S. and S. F-M gratefully acknowledge the support from the National Science Foundation under grant AST-1715876 and the SOFIA project under grant 05\_0048. J.K. acknowledges funding from the Moses Holden Studentship in support of his PhD. Financial support for SC was provided by NASA through award 08\_0186 issued by USRA. M.Tahani is supported by the Banting Fellowship (Natural Sciences and Engineering Research Council Canada) hosted at Stanford University.
We thank professor T. Miller for providing the data tables of electron density in the IRC+10\,216 shell and several helpful discussions.  We thank Prof. J. Cernicharo for pointing out the possibility that our data are dominated by line polarization.  We thank Prof. A. Lazarian and Dr. R. Sankrit for several insight- and help-full suggestions.  

The James Clerk Maxwell Telescope is operated by the East Asian Observatory on behalf of The National Astronomical Observatory of Japan; Academia Sinica Institute of Astronomy and Astrophysics; the Korea Astronomy and Space Science Institute; Center for Astronomical Mega-Science (as well as the National Key R\&D Program of China with No. 2017YFA0402700). Additional funding support is provided by the Science and Technology Facilities Council of the United Kingdom and participating universities and organizations in the United Kingdom and Canada. The authors wish to recognize and acknowledge the very significant cultural role and reverence that the summit of Maunakea has always had within the indigenous Hawaiian community.  We are most fortunate to have the opportunity to conduct observations from this mountain.
Additional funds for the construction of SCUBA-2 and POL-2 were provided by the Canada Foundation for Innovation. This research used the facilities of the Canadian Astronomy Data Centre operated by the National Research Council of Canada with the support of the Canadian Space Agency.

This work is based in part on observations made with the NASA/DLR Stratospheric Observatory for Infrared Astronomy (SOFIA). SOFIA is jointly operated by the Universities Space Research Association, Inc. (USRA), under NASA contract NNA17BF53C, and the Deutsches SOFIA Institut (DSI) under DLR contract 50 OK 0901 to the University of Stuttgart.

\end{acknowledgements}

\facility{ JCMT/POL-2, SOFIA/HAWC+
}

\software{Astropy \citep{2013A&A...558A..33A}, Matplotlib \citep{Hunter:2007}, SciPy \citep{2020SciPy-NMeth}, NumPy \citep{2020NumPy-Array}, Starlink \citep{currie2014}, SMURF \citep{chapin2011}}

\bibliography{bgbiblio_tot_new}{}

\begin{thebibliography}{}
\expandafter\ifx\csname natexlab\endcsname\relax\def\natexlab#1{#1}\fi
\providecommand{\url}[1]{\href{#1}{#1}}
\providecommand{\dodoi}[1]{doi:~\href{http://doi.org/#1}{\nolinkurl{#1}}}
\providecommand{\doeprint}[1]{\href{http://ascl.net/#1}{\nolinkurl{http://ascl.net/#1}}}
\providecommand{\doarXiv}[1]{\href{https://arxiv.org/abs/#1}{\nolinkurl{https://arxiv.org/abs/#1}}}

\bibitem[{{Abbas} {et~al.}(2006){Abbas}, {Tankosic}, {Craven}, {Spann}, {LeClair}, {West}, {Weingartner}, {Tielens}, {Nuth}, {Camata}, \& {Gerakines}}]{abbas2006}
{Abbas}, M.~M., {Tankosic}, D., {Craven}, P.~D., {et~al.} 2006, \apj, 645, 324, \dodoi{10.1086/504281}

\bibitem[{{Alves} {et~al.}(2014){Alves}, {Frau}, {Girart}, {Franco}, {Santos}, \& {Wiesemeyer}}]{alves2014}
{Alves}, F.~O., {Frau}, P., {Girart}, J.~M., {et~al.} 2014, \aap, 569, L1, \dodoi{10.1051/0004-6361/201424678}

\bibitem[{{Andersson} {et~al.}(2018){Andersson}, {Hoang}, {Lopez-Rodriguez}, {Vaillancourt}, {Sankrit}, {Lazarian}, \& {HAWC+ Instrument Team}}]{bga2018}
{Andersson}, B.~G., {Hoang}, T., {Lopez-Rodriguez}, E., {et~al.} 2018, in American Astronomical Society Meeting Abstracts, Vol. 231, American Astronomical Society Meeting Abstracts \#231, 414.04

\bibitem[{{Andersson} {et~al.}(2015){Andersson}, {Lazarian}, \& {Vaillancourt}}]{bga2015b}
{Andersson}, B.-G., {Lazarian}, A., \& {Vaillancourt}, J.~E. 2015, \araa, 53, 501, \dodoi{10.1146/annurev-astro-082214-122414}

\bibitem[{{Andersson} {et~al.}(2011){Andersson}, {Pintado}, {Potter}, {Strai{\v z}ys}, \& {Charcos-Llorens}}]{bga2011}
{Andersson}, B.-G., {Pintado}, O., {Potter}, S.~B., {Strai{\v z}ys}, V., \& {Charcos-Llorens}, M. 2011, \aap, 534, A19, \dodoi{10.1051/0004-6361/201117670}

\bibitem[{{Andersson} \& {Potter}(2007)}]{bga2007}
{Andersson}, B.-G., \& {Potter}, S.~B. 2007, \apj, 665, 369, \dodoi{10.1086/519755}

\bibitem[{{Andersson} {et~al.}(2022){Andersson}, {Lopez-Rodriguez}, {Medan}, {Soam}, {Hoang}, {Vaillancourt}, {Lazarian}, {Sandin}, {Mattsson}, \& {Tahani}}]{bga2022a}
{Andersson}, B.~G., {Lopez-Rodriguez}, E., {Medan}, I., {et~al.} 2022, \apj, 931, 80, \dodoi{10.3847/1538-4357/ac64a4}

\bibitem[{{Andrae} {et~al.}(2010){Andrae}, {Schulze-Hartung}, \& {Melchior}}]{andrae2010}
{Andrae}, R., {Schulze-Hartung}, T., \& {Melchior}, P. 2010, arXiv e-prints, arXiv:1012.3754.
\newblock \doarXiv{1012.3754}

\bibitem[{{Ashton} {et~al.}(2018){Ashton}, {Ade}, {Angil{\`e}}, {Benton}, {Devlin}, {Dober}, {Fissel}, {Fukui}, {Galitzki}, {Gandilo}, {Klein}, {Korotkov}, {Li}, {Martin}, {Matthews}, {Moncelsi}, {Nakamura}, {Netterfield}, {Novak}, {Pascale}, {Poidevin}, {Santos}, {Savini}, {Scott}, {Shariff}, {Soler}, {Thomas}, {Tucker}, {Tucker}, \& {Ward-Thompson}}]{ashton2018}
{Ashton}, P.~C., {Ade}, P. A.~R., {Angil{\`e}}, F.~E., {et~al.} 2018, \apj, 857, 10, \dodoi{10.3847/1538-4357/aab3ca}

\bibitem[{{Astropy Collaboration} {et~al.}(2013){Astropy Collaboration}, {Robitaille}, {Tollerud}, {Greenfield}, {Droettboom}, {Bray}, {Aldcroft}, {Davis}, {Ginsburg}, {Price-Whelan}, {Kerzendorf}, {Conley}, {Crighton}, {Barbary}, {Muna}, {Ferguson}, {Grollier}, {Parikh}, {Nair}, {Unther}, {Deil}, {Woillez}, {Conseil}, {Kramer}, {Turner}, {Singer}, {Fox}, {Weaver}, {Zabalza}, {Edwards}, {Azalee Bostroem}, {Burke}, {Casey}, {Crawford}, {Dencheva}, {Ely}, {Jenness}, {Labrie}, {Lim}, {Pierfederici}, {Pontzen}, {Ptak}, {Refsdal}, {Servillat}, \& {Streicher}}]{2013A&A...558A..33A}
{Astropy Collaboration}, {Robitaille}, T.~P., {Tollerud}, E.~J., {et~al.} 2013, \aap, 558, A33, \dodoi{10.1051/0004-6361/201322068}

\bibitem[{{Bastien}(2003)}]{bastien2003}
{Bastien}, P. 2003, {The circumstellar environment of AGB stars studied by polarimetry}, ed. Y.~{Nakada}, M.~{Honma}, \& M.~{Seki}, Vol. 283, 181--188, \dodoi{10.1007/978-94-010-0139-7\_35}

\bibitem[{{Bastien} {et~al.}(2011){Bastien}, {Bissonnette}, {Simon}, {Coud{\'e}}, {Ade}, {Savini}, {Pisano}, {Leclerc}, {Bernier}, {Landry}, {Houde}, {Hezareh}, {Naylor}, {Gom}, {Jenness}, {Berry}, {Johnstone}, \& {Matthews}}]{bastien2011}
{Bastien}, P., {Bissonnette}, E., {Simon}, A., {et~al.} 2011, in Astronomical Society of the Pacific Conference Series, Vol. 449, Astronomical Polarimetry 2008: Science from Small to Large Telescopes, ed. P.~{Bastien}, N.~{Manset}, D.~P. {Clemens}, \& N.~{St-Louis}, 68

\bibitem[{{Becklin} {et~al.}(1969){Becklin}, {Frogel}, {Hyland}, {Kristian}, \& {Neugebauer}}]{becklin1969}
{Becklin}, E.~E., {Frogel}, J.~A., {Hyland}, A.~R., {Kristian}, J., \& {Neugebauer}, G. 1969, \apjl, 158, L133, \dodoi{10.1086/180450}

\bibitem[{{Bevington} \& {Robinson}(1992)}]{bevington1992}
{Bevington}, P.~R., \& {Robinson}, D.~K. 1992, {Data reduction and error analysis for the physical sciences} (New York: McGraw-Hill, |c1992, 2nd ed.)

\bibitem[{{Bieging} {et~al.}(2006){Bieging}, {Schmidt}, {Smith}, \& {Oppenheimer}}]{bieging2006}
{Bieging}, J.~H., {Schmidt}, G.~D., {Smith}, P.~S., \& {Oppenheimer}, B.~D. 2006, \apj, 639, 1053, \dodoi{10.1086/499772}

\bibitem[{{Cernicharo} {et~al.}(2015){Cernicharo}, {Marcelino}, {Ag{\'u}ndez}, \& {Gu{\'e}lin}}]{cernicharo2015}
{Cernicharo}, J., {Marcelino}, N., {Ag{\'u}ndez}, M., \& {Gu{\'e}lin}, M. 2015, \aap, 575, A91, \dodoi{10.1051/0004-6361/201424565}

\bibitem[{{Cernicharo} {et~al.}(2010){Cernicharo}, {Waters}, {Decin}, {Encrenaz}, {Tielens}, {Ag{\'u}ndez}, {De Beck}, {M{\"u}ller}, {Goicoechea}, {Barlow}, {Benz}, {Crimier}, {Daniel}, {di Giorgio}, {Fich}, {Gaier}, {Garc{\'\i}a-Lario}, {de Koter}, {Khouri}, {Liseau}, {Lombaert}, {Erickson}, {Pardo}, {Pearson}, {Shipman}, {S{\'a}nchez Contreras}, \& {Teyssier}}]{cernicharo2010}
{Cernicharo}, J., {Waters}, L.~B.~F.~M., {Decin}, L., {et~al.} 2010, \aap, 521, L8, \dodoi{10.1051/0004-6361/201015150}

\bibitem[{{Chandrasekhar} \& {Fermi}(1953)}]{chandrasekhar1953}
{Chandrasekhar}, S., \& {Fermi}, E. 1953, \apj, 118, 113

\bibitem[{{Chapin} {et~al.}(2011){Chapin}, {Chapman}, {Coppin}, {Devlin}, {Dunlop}, {Greve}, {Halpern}, {Hasselfield}, {Hughes}, {Ivison}, {Marsden}, {Moncelsi}, {Netterfield}, {Pascale}, {Scott}, {Smail}, {Viero}, {Walter}, {Weiss}, \& {van der Werf}}]{chapin2011}
{Chapin}, E.~L., {Chapman}, S.~C., {Coppin}, K.~E., {et~al.} 2011, \mnras, 411, 505, \dodoi{10.1111/j.1365-2966.2010.17697.x}

\bibitem[{{Chiar} {et~al.}(2013){Chiar}, {Tielens}, {Adamson}, \& {Ricca}}]{chiar2013}
{Chiar}, J.~E., {Tielens}, A.~G.~G.~M., {Adamson}, A.~J., \& {Ricca}, A. 2013, \apj, 770, 78, \dodoi{10.1088/0004-637X/770/1/78}

\bibitem[{{Chiar} {et~al.}(2000){Chiar}, {Tielens}, {Whittet}, {Schutte}, {Boogert}, {Lutz}, {van Dishoeck}, \& {Bernstein}}]{chiar2000}
{Chiar}, J.~E., {Tielens}, A.~G.~G.~M., {Whittet}, D.~C.~B., {et~al.} 2000, \apj, 537, 749, \dodoi{10.1086/309047}

\bibitem[{{Chiar} {et~al.}(2006){Chiar}, {Adamson}, {Whittet}, {Chrysostomou}, {Hough}, {Kerr}, {Mason}, {Roche}, \& {Wright}}]{chiar2006}
{Chiar}, J.~E., {Adamson}, A.~J., {Whittet}, D.~C.~B., {et~al.} 2006, \apj, 651, 268, \dodoi{10.1086/507462}

\bibitem[{{Chuss} {et~al.}(2019){Chuss}, {Andersson}, {Bally}, {Dotson}, {Dowell}, {Guerra}, {Harper}, {Houde}, {Jones}, {Lazarian}, {Lopez Rodriguez}, {Michail}, {Morris}, {Novak}, {Siah}, {Staguhn}, {Vaillancourt}, {Volpert}, {Werner}, {Wollack}, {Benford}, {Berthoud}, {Cox}, {Crutcher}, {Dale}, {Fissel}, {Goldsmith}, {Hamilton}, {Hanany}, {Henning}, {Looney}, {Moseley}, {Santos}, {Stephens}, {Tassis}, {Trinh}, {Van Camp}, {Ward-Thompson}, \& {HAWC + Science Team}}]{chuss2019}
{Chuss}, D.~T., {Andersson}, B.~G., {Bally}, J., {et~al.} 2019, The Astrophysical Journal, 872, 187, \dodoi{10.3847/1538-4357/aafd37}

\bibitem[{{Cohen} \& {Schmidt}(1982)}]{cohen1982}
{Cohen}, M., \& {Schmidt}, G.~D. 1982, \apj, 259, 693, \dodoi{10.1086/160205}

\bibitem[{{Cookson} {et~al.}(2018){Cookson}, {Bintley}, {Li}, {Ade}, {Sudiwala}, \& {Tucker}}]{cookson2018}
{Cookson}, J.~L., {Bintley}, D., {Li}, S., {et~al.} 2018, in Society of Photo-Optical Instrumentation Engineers (SPIE) Conference Series, Vol. 10708, Millimeter, Submillimeter, and Far-Infrared Detectors and Instrumentation for Astronomy IX, ed. J.~{Zmuidzinas} \& J.-R. {Gao}, 1070839, \dodoi{10.1117/12.2313018}

\bibitem[{{Cordiner} \& {Millar}(2009)}]{cordiner2009}
{Cordiner}, M.~A., \& {Millar}, T.~J. 2009, \apj, 697, 68, \dodoi{10.1088/0004-637X/697/1/68}

\bibitem[{{Cudlip} {et~al.}(1982){Cudlip}, {Furniss}, {King}, \& {Jennings}}]{cudlip1982}
{Cudlip}, W., {Furniss}, I., {King}, K.~J., \& {Jennings}, R.~E. 1982, \mnras, 200, 1169, \dodoi{10.1093/mnras/200.4.1169}

\bibitem[{{Currie} {et~al.}(2014){Currie}, {Berry}, {Jenness}, {Gibb}, {Bell}, \& {Draper}}]{currie2014}
{Currie}, M.~J., {Berry}, D.~S., {Jenness}, T., {et~al.} 2014, in Astronomical Society of the Pacific Conference Series, Vol. 485, Astronomical Data Analysis Software and Systems XXIII, ed. N.~{Manset} \& P.~{Forshay}, 391

\bibitem[{{Davis}(1951)}]{davis1951b}
{Davis}, L. 1951, Physical Review, 81, 890, \dodoi{10.1103/PhysRev.81.890.2}

\bibitem[{{De Beck} {et~al.}(2012){De Beck}, {Lombaert}, {Ag{\'u}ndez}, {Daniel}, {Decin}, {Cernicharo}, {M{\"u}ller}, {Min}, {Royer}, {Vandenbussche}, {de Koter}, {Waters}, {Groenewegen}, {Barlow}, {Gu{\'e}lin}, {Kahane}, {Pearson}, {Encrenaz}, {Szczerba}, \& {Schmidt}}]{debeck2012}
{De Beck}, E., {Lombaert}, R., {Ag{\'u}ndez}, M., {et~al.} 2012, \aap, 539, A108, \dodoi{10.1051/0004-6361/201117635}

\bibitem[{{Decin} {et~al.}(2015){Decin}, {Richards}, {Neufeld}, {Steffen}, {Melnick}, \& {Lombaert}}]{decin2015}
{Decin}, L., {Richards}, A.~M.~S., {Neufeld}, D., {et~al.} 2015, \aap, 574, A5, \dodoi{10.1051/0004-6361/201424593}

\bibitem[{{Decin} {et~al.}(2011){Decin}, {Royer}, {Cox}, {Vandenbussche}, {Ottensamer}, {Blommaert}, {Groenewegen}, {Barlow}, {Lim}, {Kerschbaum}, {Posch}, \& {Waelkens}}]{decin2011}
{Decin}, L., {Royer}, P., {Cox}, N.~L.~J., {et~al.} 2011, \aap, 534, A1, \dodoi{10.1051/0004-6361/201117360}

\bibitem[{{Dharmawardena} {et~al.}(2018){Dharmawardena}, {Kemper}, {Scicluna}, {Wouterloot}, {Trejo}, {Srinivasan}, {Cami}, {Zijlstra}, \& {Marshall}}]{dharmawardena2018}
{Dharmawardena}, T.~E., {Kemper}, F., {Scicluna}, P., {et~al.} 2018, \mnras, 479, 536, \dodoi{10.1093/mnras/sty1422}

\bibitem[{{Dolginov} \& {Mitrofanov}(1976)}]{dolginov1976}
{Dolginov}, A.~Z., \& {Mitrofanov}, I.~G. 1976, \apss, 43, 291

\bibitem[{{Drabek} {et~al.}(2012){Drabek}, {Hatchell}, {Friberg}, {Richer}, {Graves}, {Buckle}, {Nutter}, {Johnstone}, \& {Di Francesco}}]{drabek2012}
{Drabek}, E., {Hatchell}, J., {Friberg}, P., {et~al.} 2012, \mnras, 426, 23, \dodoi{10.1111/j.1365-2966.2012.21140.x}

\bibitem[{{Draine}(2011)}]{draine2011}
{Draine}, B.~T. 2011, {Physics of the Interstellar and Intergalactic Medium} (Princeton, N.J.: Princeton University Press)

\bibitem[{{Draine} \& {Fraisse}(2009)}]{draine2009}
{Draine}, B.~T., \& {Fraisse}, A.~A. 2009, \apj, 696, 1, \dodoi{10.1088/0004-637X/696/1/1}

\bibitem[{{Draine} \& {Hensley}(2021)}]{draine2021}
{Draine}, B.~T., \& {Hensley}, B.~S. 2021, \apj, 909, 94, \dodoi{10.3847/1538-4357/abd6c6}

\bibitem[{{Draine} \& {Lazarian}(1998)}]{draine1998}
{Draine}, B.~T., \& {Lazarian}, A. 1998, \apj, 508, 157, \dodoi{10.1086/306387}

\bibitem[{{Draine} \& {Weingartner}(1996)}]{draine1996}
{Draine}, B.~T., \& {Weingartner}, J.~C. 1996, \apj, 470, 551, \dodoi{10.1086/177887}

\bibitem[{{Duthu} {et~al.}(2017){Duthu}, {Herpin}, {Wiesemeyer}, {Baudry}, {L{\`e}bre}, \& {Paubert}}]{duthu2017}
{Duthu}, A., {Herpin}, F., {Wiesemeyer}, H., {et~al.} 2017, \aap, 604, A12, \dodoi{10.1051/0004-6361/201730485}

\bibitem[{{Dyck} {et~al.}(1971){Dyck}, {Forbes}, \& {Shawl}}]{dyck1971}
{Dyck}, H.~M., {Forbes}, F.~F., \& {Shawl}, S.~J. 1971, \aj, 76, 901, \dodoi{10.1086/111199}

\bibitem[{{Gail} \& {Sedlmayr}(2013)}]{gail2013}
{Gail}, H.-P., \& {Sedlmayr}, E. 2013, {Physics and Chemistry of Circumstellar Dust Shells}

\bibitem[{{Girart} {et~al.}(2012){Girart}, {Patel}, {Vlemmings}, \& {Rao}}]{girart2012}
{Girart}, J.~M., {Patel}, N., {Vlemmings}, W.~H.~T., \& {Rao}, R. 2012, \apjl, 751, L20, \dodoi{10.1088/2041-8205/751/1/L20}

\bibitem[{{Girart} {et~al.}(2006){Girart}, {Rao}, \& {Marrone}}]{girart2006}
{Girart}, J.~M., {Rao}, R., \& {Marrone}, D.~P. 2006, Science, 313, 812, \dodoi{10.1126/science.1129093}

\bibitem[{{Gold}(1952)}]{gold1952}
{Gold}, T. 1952, \mnras, 112, 215

\bibitem[{{Goldreich} \& {Kylafis}(1981)}]{goldreich1981}
{Goldreich}, P., \& {Kylafis}, N.~D. 1981, \apjl, 243, L75, \dodoi{10.1086/183446}

\bibitem[{{Goldreich} \& {Kylafis}(1982)}]{goldreich1982}
---. 1982, \apj, 253, 606, \dodoi{10.1086/159663}

\bibitem[{{Goodman} {et~al.}(1995){Goodman}, {Jones}, {Lada}, \& {Myers}}]{goodman1995}
{Goodman}, A.~A., {Jones}, T.~J., {Lada}, E.~A., \& {Myers}, P.~C. 1995, \apj, 448, 748

\bibitem[{{Groenewegen}(1997)}]{groenewegen1997}
{Groenewegen}, M.~A.~T. 1997, \aap, 317, 503

\bibitem[{{Groenewegen} {et~al.}(2012){Groenewegen}, {Barlow}, {Blommaert}, {Cernicharo}, {Decin}, {Gomez}, {Hargrave}, {Kerschbaum}, {Ladjal}, {Lim}, {Matsuura}, {Olofsson}, {Sibthorpe}, {Swinyard}, {Ueta}, \& {Yates}}]{groenewegen2012}
{Groenewegen}, M.~A.~T., {Barlow}, M.~J., {Blommaert}, J.~A.~D.~L., {et~al.} 2012, Astronomy and Astrophysics, 543, L8, \dodoi{10.1051/0004-6361/201219604}

\bibitem[{{Gu{\'e}lin} {et~al.}(2018){Gu{\'e}lin}, {Patel}, {Bremer}, {Cernicharo}, {Castro-Carrizo}, {Pety}, {Fonfr{\'\i}a}, {Ag{\'u}ndez}, {Santander-Garc{\'\i}a}, {Quintana-Lacaci}, {Velilla Prieto}, {Blundell}, \& {Thaddeus}}]{guelin2018}
{Gu{\'e}lin}, M., {Patel}, N.~A., {Bremer}, M., {et~al.} 2018, \aap, 610, A4, \dodoi{10.1051/0004-6361/201731619}

\bibitem[{{Guzman-Ramirez} {et~al.}(2015){Guzman-Ramirez}, {Lagadec}, {Wesson}, {Zijlstra}, {Muller}, {Jones}, {Boffin}, {Sloan}, {Redman}, {Smette}, {Karakas}, \& {Nyman}}]{guzman-ramirez2015}
{Guzman-Ramirez}, L., {Lagadec}, E., {Wesson}, R., {et~al.} 2015, \mnras, 451, L1, \dodoi{10.1093/mnrasl/slv055}

\bibitem[{{Hall}(1949)}]{hall1949}
{Hall}, J.~S. 1949, Science, 109, 166

\bibitem[{Harris {et~al.}(2020)Harris, Millman, van~der Walt, Gommers, Virtanen, Cournapeau, Wieser, Taylor, Berg, Smith, Kern, Picus, Hoyer, van Kerkwijk, Brett, Haldane, Fernández~del Río, Wiebe, Peterson, Gérard-Marchant, Sheppard, Reddy, Weckesser, Abbasi, Gohlke, \& Oliphant}]{2020NumPy-Array}
Harris, C.~R., Millman, K.~J., van~der Walt, S.~J., {et~al.} 2020, Nature, 585, 357–362, \dodoi{10.1038/s41586-020-2649-2}

\bibitem[{{Hildebrand} {et~al.}(2000){Hildebrand}, {Davidson}, {Dotson}, {Dowell}, {Novak}, \& {Vaillancourt}}]{hildebrand2000}
{Hildebrand}, R.~H., {Davidson}, J.~A., {Dotson}, J.~L., {et~al.} 2000, \pasp, 112, 1215, \dodoi{10.1086/316613}

\bibitem[{{Hiltner}(1949{\natexlab{a}})}]{hiltner1949a}
{Hiltner}, W.~A. 1949{\natexlab{a}}, Science, 109, 165

\bibitem[{{Hiltner}(1949{\natexlab{b}})}]{hiltner1949b}
---. 1949{\natexlab{b}}, \apj, 109, 471

\bibitem[{{Hoang} \& {Lazarian}(2009)}]{hoang2009a}
{Hoang}, T., \& {Lazarian}, A. 2009, \apj, 697, 1316, \dodoi{10.1088/0004-637X/697/2/1316}

\bibitem[{{Hoang} \& {Lee}(2020)}]{hoang2020}
{Hoang}, T., \& {Lee}, H. 2020, \apj, 896, 144, \dodoi{10.3847/1538-4357/ab9609}

\bibitem[{{Hoang} {et~al.}(2019){Hoang}, {Tram}, {Lee}, \& {Ahn}}]{hoang2019}
{Hoang}, T., {Tram}, L.~N., {Lee}, H., \& {Ahn}, S.-H. 2019, Nature Astronomy, 3, 766, \dodoi{10.1038/s41550-019-0763-6}

\bibitem[{{Hoang} {et~al.}(2021){Hoang}, {Tram}, {Lee}, {Diep}, \& {Ngoc}}]{hoang2021}
{Hoang}, T., {Tram}, L.~N., {Lee}, H., {Diep}, P.~N., \& {Ngoc}, N.~B. 2021, \apj, 908, 218, \dodoi{10.3847/1538-4357/abd54f}

\bibitem[{{Holland} {et~al.}(2013){Holland}, {Bintley}, {Chapin}, {Chrysostomou}, {Davis}, {Dempsey}, {Duncan}, {Fich}, {Friberg}, {Halpern}, {Irwin}, {Jenness}, {Kelly}, {MacIntosh}, {Robson}, {Scott}, {Ade}, {Atad-Ettedgui}, {Berry}, {Craig}, {Gao}, {Gibb}, {Hilton}, {Hollister}, {Kycia}, {Lunney}, {McGregor}, {Montgomery}, {Parkes}, {Tilanus}, {Ullom}, {Walther}, {Walton}, {Woodcraft}, {Amiri}, {Atkinson}, {Burger}, {Chuter}, {Coulson}, {Doriese}, {Dunare}, {Economou}, {Niemack}, {Parsons}, {Reintsema}, {Sibthorpe}, {Smail}, {Sudiwala}, \& {Thomas}}]{holland2013}
{Holland}, W.~S., {Bintley}, D., {Chapin}, E.~L., {et~al.} 2013, \mnras, 430, 2513, \dodoi{10.1093/mnras/sts612}

\bibitem[{{Houde} {et~al.}(2009){Houde}, {Vaillancourt}, {Hildebrand}, {Chitsazzadeh}, \& {Kirby}}]{houde2009}
{Houde}, M., {Vaillancourt}, J.~E., {Hildebrand}, R.~H., {Chitsazzadeh}, S., \& {Kirby}, L. 2009, \apj, 706, 1504, \dodoi{10.1088/0004-637X/706/2/1504}

\bibitem[{Hunter(2007)}]{Hunter:2007}
Hunter, J.~D. 2007, Computing in Science \& Engineering, 9, 90, \dodoi{10.1109/MCSE.2007.55}

\bibitem[{{Ib{\'a}{\~n}ez-Mej{\'\i}a} {et~al.}(2019){Ib{\'a}{\~n}ez-Mej{\'\i}a}, {Walch}, {Ivlev}, {Clarke}, {Caselli}, \& {Joshi}}]{ibanez-mejia2019}
{Ib{\'a}{\~n}ez-Mej{\'\i}a}, J.~C., {Walch}, S., {Ivlev}, A.~V., {et~al.} 2019, \mnras, 485, 1220, \dodoi{10.1093/mnras/stz207}

\bibitem[{{Ivezi{\'c}} \& {Elitzur}(1996{\natexlab{a}})}]{ivezic1996a}
{Ivezi{\'c}}, Z., \& {Elitzur}, M. 1996{\natexlab{a}}, \mnras, 279, 1019, \dodoi{10.1093/mnras/279.3.1019}

\bibitem[{{Ivezi{\'c}} \& {Elitzur}(1996{\natexlab{b}})}]{ivezic1996b}
---. 1996{\natexlab{b}}, \mnras, 279, 1011, \dodoi{10.1093/mnras/279.3.1011}

\bibitem[{{Ivezic} {et~al.}(1999){Ivezic}, {Nenkova}, \& {Elitzur}}]{ivezic1999}
{Ivezic}, Z., {Nenkova}, M., \& {Elitzur}, M. 1999, {DUSTY: Radiation transport in a dusty environment}.
\newblock \doeprint{9911.001}

\bibitem[{{Ivlev} {et~al.}(2015){Ivlev}, {Padovani}, {Galli}, \& {Caselli}}]{ivlev2015}
{Ivlev}, A.~V., {Padovani}, M., {Galli}, D., \& {Caselli}, P. 2015, \apj, 812, 135, \dodoi{10.1088/0004-637X/812/2/135}

\bibitem[{{Jeffers} {et~al.}(2014){Jeffers}, {Min}, {Waters}, {Canovas}, {Pols}, {Rodenhuis}, {de Juan Ovelar}, {Keller}, \& {Decin}}]{jeffers2014}
{Jeffers}, S.~V., {Min}, M., {Waters}, L.~B.~F.~M., {et~al.} 2014, \aap, 572, A3, \dodoi{10.1051/0004-6361/201423463}

\bibitem[{{Jones} \& {Spitzer}(1967)}]{jones1967}
{Jones}, R.~V., \& {Spitzer}, Jr., L. 1967, \apj, 147, 943, \dodoi{10.1086/149086}

\bibitem[{{Jones} {et~al.}(2015){Jones}, {Bagley}, {Krejny}, {Andersson}, \& {Bastien}}]{jones2015}
{Jones}, T.~J., {Bagley}, M., {Krejny}, M., {Andersson}, B.-G., \& {Bastien}, P. 2015, \aj, 149, 31, \dodoi{10.1088/0004-6256/149/1/31}

\bibitem[{{Kahane} {et~al.}(1997){Kahane}, {Viard}, {M{\'e}nard}, {Bastien}, \& {Manset}}]{kahane1997}
{Kahane}, C., {Viard}, E., {M{\'e}nard}, F., {Bastien}, P., \& {Manset}, N. 1997, \apss, 251, 223, \dodoi{10.1023/A:1000773718832}

\bibitem[{{Kastner} \& {Weintraub}(1994)}]{kastner1994}
{Kastner}, J.~H., \& {Weintraub}, D.~A. 1994, \apj, 434, 719, \dodoi{10.1086/174774}

\bibitem[{{Kastner} \& {Weintraub}(1996)}]{kastner1996}
{Kastner}, J.~H., \& {Weintraub}, D.~A. 1996, in Astronomical Society of the Pacific Conference Series, Vol.~97, Polarimetry of the Interstellar Medium, ed. W.~G. {Roberge} \& D.~C.~B. {Whittet}, 212

\bibitem[{{Kataoka} {et~al.}(2017){Kataoka}, {Tsukagoshi}, {Pohl}, {Muto}, {Nagai}, {Stephens}, {Tomisaka}, \& {Momose}}]{kataoka2017}
{Kataoka}, A., {Tsukagoshi}, T., {Pohl}, A., {et~al.} 2017, The Astrophysical Journal, 844, L5, \dodoi{10.3847/2041-8213/aa7e33}

\bibitem[{{Khouri} {et~al.}(2020){Khouri}, {Vlemmings}, {Paladini}, {Ginski}, {Lagadec}, {Maercker}, {Kervella}, {De Beck}, {Decin}, {de Koter}, \& {Waters}}]{Khouri2020}
{Khouri}, T., {Vlemmings}, W.~H.~T., {Paladini}, C., {et~al.} 2020, \aap, 635, A200, \dodoi{10.1051/0004-6361/201834618}

\bibitem[{{Kim} {et~al.}(2015){Kim}, {Lee}, {Mauron}, \& {Chu}}]{kim2015}
{Kim}, H., {Lee}, H.-G., {Mauron}, N., \& {Chu}, Y.-H. 2015, \apjl, 804, L10, \dodoi{10.1088/2041-8205/804/1/L10}

\bibitem[{{Kim} {et~al.}(2021){Kim}, {Lee}, {Ohyama}, {Kim}, {Scicluna}, {Chu}, {Mauron}, \& {Ueta}}]{kim2021}
{Kim}, H., {Lee}, H.-G., {Ohyama}, Y., {et~al.} 2021, \apj, 914, 35, \dodoi{10.3847/1538-4357/abf6cc}

\bibitem[{{Kim} \& {Martin}(1995)}]{kim1995}
{Kim}, S.-H., \& {Martin}, P.~G. 1995, \apj, 444, 293, \dodoi{10.1086/175604}

\bibitem[{{King} {et~al.}(2018){King}, {Fissel}, {Chen}, \& {Li}}]{king2018}
{King}, P.~K., {Fissel}, L.~M., {Chen}, C.-Y., \& {Li}, Z.-Y. 2018, \mnras, 474, 5122, \dodoi{10.1093/mnras/stx3096}

\bibitem[{{Kwon} {et~al.}(2018){Kwon}, {Doi}, {Tamura}, {Matsumura}, {Pattle}, {Berry}, {Sadavoy}, {Matthews}, {Ward-Thompson}, {Hasegawa}, {Furuya}, {Pon}, {Di Francesco}, {Arzoumanian}, {Hayashi}, {Kawabata}, {Onaka}, {Choi}, {Kang}, {Hoang}, {Lee}, {Lee}, {Liu}, {Liu}, {Inutsuka}, {Eswaraiah}, {Bastien}, {Kwon}, {Lai}, {Qiu}, {Coud{\'e}}, {Franzmann}, {Friberg}, {Graves}, {Greaves}, {Houde}, {Johnstone}, {Kirk}, {Koch}, {Li}, {Parsons}, {Rao}, {Rawlings}, {Shinnaga}, {van Loo}, {Aso}, {Byun}, {Chen}, {Chen}, {Chen}, {Ching}, {Cho}, {Chrysostomou}, {Chung}, {Drabek-Maunder}, {Eyres}, {Fiege}, {Friesen}, {Fuller}, {Gledhill}, {Griffin}, {Gu}, {Hatchell}, {Holland}, {Inoue}, {Iwasaki}, {Jeong}, {Kang}, {Kang}, {Kemper}, {Kim}, {Kim}, {Kim}, {Kim}, {Kim}, {Kim}, {Lacaille}, {Lee}, {Li}, {Li}, {Liu}, {Liu}, {Lyo}, {Mairs}, {Moriarty-Schieven}, {Nakamura}, {Nakanishi}, {Ohashi}, {Peretto}, {Pyo}, {Qian}, {Retter}, {Richer}, {Rigby}, {Robitaille}, {Savini}, {Scaife}, {Soam}, {Tang}, {Tomisaka}, {Wang}, {Wang},
  {Whitworth}, {Yen}, {Yoo}, {Yuan}, {Zhang}, {Zhang}, {Zhou}, {Zhu}, {Andr{\'e}}, {Dowell}, {Falle}, {Tsukamoto}, {Nakagawa}, {Kanamori}, {Kataoka}, {Kobayashi}, {Nagata}, {Saito}, {Seta}, \& {Zenko}}]{kwon2018}
{Kwon}, J., {Doi}, Y., {Tamura}, M., {et~al.} 2018, \apj, 859, 4, \dodoi{10.3847/1538-4357/aabd82}

\bibitem[{{Ladjal} {et~al.}(2010){Ladjal}, {Barlow}, {Groenewegen}, {Ueta}, {Blommaert}, {Cohen}, {Decin}, {De Meester}, {Exter}, {Gear}, {Gomez}, {Hargrave}, {Huygen}, {Ivison}, {Jean}, {Kerschbaum}, {Leeks}, {Lim}, {Olofsson}, {Polehampton}, {Posch}, {Regibo}, {Royer}, {Sibthorpe}, {Swinyard}, {Vandenbussche}, {Waelkens}, \& {Wesson}}]{ladjal2010}
{Ladjal}, D., {Barlow}, M.~J., {Groenewegen}, M.~A.~T., {et~al.} 2010, \aap, 518, L141, \dodoi{10.1051/0004-6361/201014658}

\bibitem[{{Lattanzio} \& {Wood}(2004)}]{lattanzio2004}
{Lattanzio}, J.~C., \& {Wood}, P.~R. 2004, {Evolution, Nucleosynthesis, and Pulsation of AGB Stars}, 23--104, \dodoi{10.1007/978-1-4757-3876-6_2}

\bibitem[{{Lazarian}(2020)}]{lazarian2020}
{Lazarian}, A. 2020, \apj, 902, 97, \dodoi{10.3847/1538-4357/abb1b4}

\bibitem[{{Lazarian} \& {Draine}(1999)}]{lazarian1999b}
{Lazarian}, A., \& {Draine}, B.~T. 1999, \apjl, 516, L37, \dodoi{10.1086/311986}

\bibitem[{{Lazarian} \& {Hoang}(2007)}]{lazarian2007a}
{Lazarian}, A., \& {Hoang}, T. 2007, \mnras, 378, 910, \dodoi{10.1111/j.1365-2966.2007.11817.x}

\bibitem[{{Lazarian} \& {Hoang}(2021)}]{lazarian2021}
---. 2021, \apj, 908, 12, \dodoi{10.3847/1538-4357/abd02c}

\bibitem[{{Lazarian} {et~al.}(2022){Lazarian}, {Yuen}, \& {Pogosyan}}]{lazarian2022}
{Lazarian}, A., {Yuen}, K.~H., \& {Pogosyan}, D. 2022, \apj, 935, 77, \dodoi{10.3847/1538-4357/ac6877}

\bibitem[{{Le{\~a}o} {et~al.}(2006){Le{\~a}o}, {de Laverny}, {M{\'e}karnia}, {de Medeiros}, \& {Vandame}}]{leao2006}
{Le{\~a}o}, I.~C., {de Laverny}, P., {M{\'e}karnia}, D., {de Medeiros}, J.~R., \& {Vandame}, B. 2006, \aap, 455, 187, \dodoi{10.1051/0004-6361:20054577}

\bibitem[{{Li} {et~al.}(2014){Li}, {Millar}, {Walsh}, {Heays}, \& {van Dishoeck}}]{li2014}
{Li}, X., {Millar}, T.~J., {Walsh}, C., {Heays}, A.~N., \& {van Dishoeck}, E.~F. 2014, \aap, 568, A111, \dodoi{10.1051/0004-6361/201424076}

\bibitem[{{Liu} {et~al.}(2019){Liu}, {Qiu}, {Berry}, {Di Francesco}, {Bastien}, {Koch}, {Furuya}, {Kim}, {Coud{\'e}}, {Lee}, {Soam}, {Eswaraiah}, {Li}, {Hwang}, {Lyo}, {Pattle}, {Hasegawa}, {Kwon}, {Lai}, {Ward-Thompson}, {Ching}, {Chen}, {Gu}, {Li}, {Li}, {Liu}, {Qian}, {Wang}, {Yuan}, {Zhang}, {Zhang}, {Zhang}, {Zhou}, {Zhu}, {Andr{\'e}}, {Arzoumanian}, {Aso}, {Byun}, {Chen}, {Chen}, {Chen}, {Cho}, {Choi}, {Chrysostomou}, {Chung}, {Doi}, {Drabek-Maunder}, {Dowell}, {Eyres}, {Falle}, {Fanciullo}, {Fiege}, {Franzmann}, {Friberg}, {Friesen}, {Fuller}, {Gledhill}, {Graves}, {Greaves}, {Griffin}, {Han}, {Hatchell}, {Hayashi}, {Hoang}, {Holland}, {Houde}, {Inoue}, {Inutsuka}, {Iwasaki}, {Jeong}, {Johnstone}, {Kanamori}, {Kang}, {Kang}, {Kang}, {Kataoka}, {Kawabata}, {Kemper}, {Kim}, {Kim}, {Kim}, {Kim}, {Kim}, {Kirk}, {Kobayashi}, {Kusune}, {Kwon}, {Lacaille}, {Lee}, {Lee}, {Lee}, {Lee}, {Liu}, {Liu}, {van Loo}, {Mairs}, {Matsumura}, {Matthews}, {Moriarty-Schieven}, {Nagata}, {Nakamura}, {Nakanishi}, {Ohashi},
  {Onaka}, {Parker}, {Parsons}, {Pascale}, {Peretto}, {Pon}, {Pyo}, {Rao}, {Rawlings}, {Retter}, {Richer}, {Rigby}, {Robitaille}, {Sadavoy}, {Saito}, {Savini}, {Scaife}, {Seta}, {Shinnaga}, {Tamura}, {Tang}, {Tomisaka}, {Tsukamoto}, {Wang}, {Whitworth}, {Yen}, {Yoo}, \& {Zenko}}]{liu2019b}
{Liu}, J., {Qiu}, K., {Berry}, D., {et~al.} 2019, \apj, 877, 43, \dodoi{10.3847/1538-4357/ab0958}

\bibitem[{{Lopez-Rodriguez} {et~al.}(2020){Lopez-Rodriguez}, {Alonso-Herrero}, {Garc{\'\i}a-Burillo}, {Gordon}, {Ichikawa}, {Imanishi}, {Kameno}, {Levenson}, {Nikutta}, \& {Packham}}]{lopezrodriguez2020}
{Lopez-Rodriguez}, E., {Alonso-Herrero}, A., {Garc{\'\i}a-Burillo}, S., {et~al.} 2020, \apj, 893, 33, \dodoi{10.3847/1538-4357/ab8013}

\bibitem[{{Lupton}(1993)}]{lupton1993}
{Lupton}, R. 1993, {Statistics in theory and practice} (Princeton, N.J.: Princeton University Press).
\newblock \url{http://adsabs.harvard.edu/cgi-bin/nph-bib_query?bibcode=1993stp..book.....L&db_key=AST}

\bibitem[{{Mathis}(1986)}]{mathis1986}
{Mathis}, J.~S. 1986, \apj, 308, 281, \dodoi{10.1086/164499}

\bibitem[{{Matthews} {et~al.}(2009){Matthews}, {McPhee}, {Fissel}, \& {Curran}}]{matthews2009}
{Matthews}, B.~C., {McPhee}, C.~A., {Fissel}, L.~M., \& {Curran}, R.~L. 2009, \apjs, 182, 143, \dodoi{10.1088/0067-0049/182/1/143}

\bibitem[{{Mauron} \& {Huggins}(1999)}]{mauron1999}
{Mauron}, N., \& {Huggins}, P.~J. 1999, \aap, 349, 203

\bibitem[{{Mauron} \& {Huggins}(2000)}]{mauron2000}
---. 2000, \aap, 359, 707

\bibitem[{{Mauron} \& {Huggins}(2010)}]{mauron2010}
---. 2010, \aap, 513, A31, \dodoi{10.1051/0004-6361/200913970}

\bibitem[{{Medan} \& {Andersson}(2019)}]{medan2019}
{Medan}, I., \& {Andersson}, B.-G. 2019, \apj, 873, 87, \dodoi{10.3847/1538-4357/ab063c}

\bibitem[{{Milam} {et~al.}(2009){Milam}, {Woolf}, \& {Ziurys}}]{milam2009}
{Milam}, S.~N., {Woolf}, N.~J., \& {Ziurys}, L.~M. 2009, \apj, 690, 837, \dodoi{10.1088/0004-637X/690/1/837}

\bibitem[{{Miller}(1970)}]{miller1970}
{Miller}, J.~S. 1970, \apjl, 161, L95, \dodoi{10.1086/180578}

\bibitem[{{Min} {et~al.}(2009){Min}, {Dullemond}, {Dominik}, {de Koter}, \& {Hovenier}}]{min2009}
{Min}, M., {Dullemond}, C.~P., {Dominik}, C., {de Koter}, A., \& {Hovenier}, J.~W. 2009, \aap, 497, 155, \dodoi{10.1051/0004-6361/200811470}

\bibitem[{{Montarg{\`e}s} {et~al.}(2023){Montarg{\`e}s}, {Cannon}, {de Koter}, {Khouri}, {Lagadec}, {Kervella}, {Decin}, {McDonald}, {Homan}, {Waters}, {Sahai}, {Gottlieb}, {Malfait}, {Maes}, {Pimpanuwat}, {Jeste}, {Danilovich}, {De Ceuster}, {Van de Sande}, {Gobrecht}, {Wallstr{\"o}m}, {Wong}, {El Mellah}, {Bolte}, {Herpin}, {Richards}, {Baudry}, {Etoka}, {Gray}, {Millar}, {Menten}, {M{\"u}ller}, {Plane}, {Yates}, \& {Zijlstra}}]{montarges2023}
{Montarg{\`e}s}, M., {Cannon}, E., {de Koter}, A., {et~al.} 2023, \aap, 671, A96, \dodoi{10.1051/0004-6361/202245398}

\bibitem[{{Murakawa} {et~al.}(2005){Murakawa}, {Suto}, {Oya}, {Yates}, {Ueta}, \& {Meixner}}]{murakawa2005}
{Murakawa}, K., {Suto}, H., {Oya}, S., {et~al.} 2005, \aap, 436, 601, \dodoi{10.1051/0004-6361:20042477}

\bibitem[{{Naghizadeh-Khouei} \& {Clarke}(1993)}]{naghizadeh1993}
{Naghizadeh-Khouei}, J., \& {Clarke}, D. 1993, \aap, 274, 968

\bibitem[{{Olofsson}(2004)}]{olofsson2004}
{Olofsson}, H. 2004, {Circumstellar Envelopes}, 325--410, \dodoi{10.1007/978-1-4757-3876-6_7}

\bibitem[{{Pattle} {et~al.}(2021){Pattle}, {Lai}, {Wright}, {Coud{\'e}}, {Plambeck}, {Hoang}, {Tang}, {Bastien}, {Eswaraiah}, {Furuya}, {Hwang}, {Inutsuka}, {Kim}, {Kirchschlager}, {Kwon}, {Lee}, {Liu}, {Lyo}, {Ohashi}, {Rawlings}, {Tahani}, {Tamura}, {Soam}, {Wang}, \& {Ward-Thompson}}]{pattle2021}
{Pattle}, K., {Lai}, S.-P., {Wright}, M., {et~al.} 2021, \mnras, 503, 3414, \dodoi{10.1093/mnras/stab608}

\bibitem[{{Purcell}(1979)}]{purcell1979}
{Purcell}, E.~M. 1979, \apj, 231, 404, \dodoi{10.1086/157204}

\bibitem[{{Reach} {et~al.}(2022){Reach}, {Ruaud}, {Wiesemeyer}, {Riquelme}, {Tram}, {Cernicharo}, {Smith}, \& {Chambers}}]{reach2022}
{Reach}, W.~T., {Ruaud}, M., {Wiesemeyer}, H., {et~al.} 2022, \apj, 926, 69, \dodoi{10.3847/1538-4357/ac4162}

\bibitem[{{Sahai} \& {Chronopoulos}(2010)}]{sahai2010}
{Sahai}, R., \& {Chronopoulos}, C.~K. 2010, \apjl, 711, L53, \dodoi{10.1088/2041-8205/711/2/L53}

\bibitem[{{Santos} {et~al.}(2019){Santos}, {Chuss}, {Dowell}, {Houde}, {Looney}, {Lopez Rodriguez}, {Novak}, {Ward-Thompson}, {Berthoud}, {Dale}, {Guerra}, {Hamilton}, {Hanany}, {Harper}, {Henning}, {Jones}, {Lazarian}, {Michail}, {Morris}, {Staguhn}, {Stephens}, {Tassis}, {Trinh}, {Van Camp}, {Volpert}, \& {Wollack}}]{santos2019}
{Santos}, F.~P., {Chuss}, D.~T., {Dowell}, C.~D., {et~al.} 2019, \apj, 882, 113, \dodoi{10.3847/1538-4357/ab3407}

\bibitem[{{Siebert} {et~al.}(2022){Siebert}, {Van de Sande}, {Millar}, \& {Remijan}}]{siebert2022}
{Siebert}, M.~A., {Van de Sande}, M., {Millar}, T.~J., \& {Remijan}, A.~J. 2022, \apj, 941, 90, \dodoi{10.3847/1538-4357/ac9e52}

\bibitem[{{Sloan} {et~al.}(1998){Sloan}, {Little-Marenin}, \& {Price}}]{sloan1998}
{Sloan}, G.~C., {Little-Marenin}, I.~R., \& {Price}, S.~D. 1998, \aj, 115, 809, \dodoi{10.1086/300205}

\bibitem[{{Smith} {et~al.}(2000){Smith}, {Wright}, {Aitken}, {Roche}, \& {Hough}}]{smith2000}
{Smith}, C.~H., {Wright}, C.~M., {Aitken}, D.~K., {Roche}, P.~F., \& {Hough}, J.~H. 2000, \mnras, 312, 327, \dodoi{10.1046/j.1365-8711.2000.03158.x}

\bibitem[{{Soam} {et~al.}(2021){Soam}, {Andersson}, {Strai{\v{z}}ys}, {Caputo}, {Kazlauskas}, {Boyle}, {Janusz}, {Zdanavi{\v{c}}ius}, \& {Acosta-Pulido}}]{soam2021a}
{Soam}, A., {Andersson}, B.~G., {Strai{\v{z}}ys}, V., {et~al.} 2021, \aj, 161, 149, \dodoi{10.3847/1538-3881/abdd3b}

\bibitem[{{Szymczak} {et~al.}(2001){Szymczak}, {Cohen}, \& {Richards}}]{szymczak2001}
{Szymczak}, M., {Cohen}, R.~J., \& {Richards}, A.~M.~S. 2001, \aap, 371, 1012, \dodoi{10.1051/0004-6361:20010451}

\bibitem[{{Tabernero} {et~al.}(2021){Tabernero}, {Dorda}, {Negueruela}, \& {Marfil}}]{tabernero2021}
{Tabernero}, H.~M., {Dorda}, R., {Negueruela}, I., \& {Marfil}, E. 2021, \aap, 646, A98, \dodoi{10.1051/0004-6361/202039236}

\bibitem[{{Tahani} {et~al.}(2023){Tahani}, {Bastien}, {Furuya}, {Pattle}, {Johnstone}, {Arzoumanian}, {Doi}, {Hasegawa}, {Inutsuka}, {Coud{\'e}}, {Fissel}, {Chen}, {Poidevin}, {Sadavoy}, {Friesen}, {Koch}, {Di Francesco}, {Moriarty-Schieven}, {Chen}, {Chung}, {Eswaraiah}, {Fanciullo}, {Gledhill}, {Le Gouellec}, {Hoang}, {Hwang}, {Kang}, {Kim}, {Kirchschlager}, {Kwon}, {Lee}, {Liu}, {Onaka}, {Rawlings}, {Soam}, {Tamura}, {Tang}, {Tomisaka}, {Whitworth}, {Kwon}, {Hoang}, {Redman}, {Berry}, {Ching}, {Wang}, {Lai}, {Qiu}, {Ward-Thompson}, {Houde}, {Byun}, {Chen}, {Chen}, {Cho}, {Choi}, {Choi}, {Chrysostomou}, {Diep}, {Duan}, {Fiege}, {Franzmann}, {Friberg}, {Fuller}, {Graves}, {Greaves}, {Griffin}, {Gu}, {Han}, {Hatchell}, {Hayashi}, {Hull}, {Inoue}, {Iwasaki}, {Jeong}, {Kanamori}, {Kang}, {Kang}, {Kataoka}, {Kawabata}, {Kemper}, {Kim}, {Kim}, {Kim}, {Kim}, {Kim}, {Kirk}, {Kobayashi}, {Konyves}, {Kusune}, {Lacaille}, {Law}, {Lee}, {Lee}, {Lee}, {Lee}, {Lee}, {Li}, {Li}, {Li}, {Liu}, {Liu}, {Liu}, {de Looze},
  {Lyo}, {Mairs}, {Matsumura}, {Matthews}, {Nagata}, {Nakamura}, {Nakanishi}, {Ohashi}, {Park}, {Parsons}, {Peretto}, {Pyo}, {Qian}, {Rao}, {Retter}, {Richer}, {Rigby}, {Saito}, {Savini}, {Scaife}, {Seta}, {Shimajiri}, {Shinnaga}, {Tang}, {Tsukamoto}, {Viti}, {Wang}, {Yen}, {Yoo}, {Yuan}, {Yun}, {Zenko}, {Zhang}, {Zhang}, {Zhang}, {Zhou}, {Zhu}, {Andr{\'e}}, {Dowell}, {Eyres}, {Falle}, {van Loo}, \& {Robitaille}}]{tahani2023}
{Tahani}, M., {Bastien}, P., {Furuya}, R.~S., {et~al.} 2023, \apj, 944, 139, \dodoi{10.3847/1538-4357/acac81}

\bibitem[{{Trammell} {et~al.}(1994){Trammell}, {Dinerstein}, \& {Goodrich}}]{trammell1994}
{Trammell}, S.~R., {Dinerstein}, H.~L., \& {Goodrich}, R.~W. 1994, \aj, 108, 984, \dodoi{10.1086/117128}

\bibitem[{{Vaillancourt} \& {Andersson}(2015)}]{vaillancourt2015}
{Vaillancourt}, J.~E., \& {Andersson}, B.-G. 2015, \apjl, 812, L7, \dodoi{10.1088/2041-8205/812/1/L7}

\bibitem[{{Vaillancourt} {et~al.}(2020){Vaillancourt}, {Andersson}, {Clemens}, {Piirola}, {Hoang}, {Becklin}, \& {Caputo}}]{vaillancourt2020}
{Vaillancourt}, J.~E., {Andersson}, B.-G., {Clemens}, D.~P., {et~al.} 2020, \apj, 905, 157, \dodoi{10.3847/1538-4357/abc6b0}

\bibitem[{{Vaillancourt} {et~al.}(2008){Vaillancourt}, {Dowell}, {Hildebrand}, {Kirby}, {Krejny}, {Li}, {Novak}, {Houde}, {Shinnaga}, \& {Attard}}]{vaillancourt2008}
{Vaillancourt}, J.~E., {Dowell}, C.~D., {Hildebrand}, R.~H., {et~al.} 2008, \apjl, 679, L25, \dodoi{10.1086/589152}

\bibitem[{Virtanen {et~al.}(2020)Virtanen, Gommers, Oliphant, Haberland, Reddy, Cournapeau, Burovski, Peterson, Weckesser, Bright, {van der Walt}, Brett, Wilson, Millman, Mayorov, Nelson, Jones, Kern, Larson, Carey, Polat, Feng, Moore, {VanderPlas}, Laxalde, Perktold, Cimrman, Henriksen, Quintero, Harris, Archibald, Ribeiro, Pedregosa, {van Mulbregt}, \& {SciPy 1.0 Contributors}}]{2020SciPy-NMeth}
Virtanen, P., Gommers, R., Oliphant, T.~E., {et~al.} 2020, Nature Methods, 17, 261, \dodoi{10.1038/s41592-019-0686-2}

\bibitem[{{Vlemmings}(2012)}]{vlemmings2012}
{Vlemmings}, W.~H.~T. 2012, in Cosmic Masers - from OH to H0, ed. R.~S. {Booth}, W.~H.~T. {Vlemmings}, \& E.~M.~L. {Humphreys}, Vol. 287, 31--40, \dodoi{10.1017/S1743921312006606}

\bibitem[{{Vlemmings} {et~al.}(2005){Vlemmings}, {van Langevelde}, \& {Diamond}}]{vlemmings2005}
{Vlemmings}, W.~H.~T., {van Langevelde}, H.~J., \& {Diamond}, P.~J. 2005, \aap, 434, 1029, \dodoi{10.1051/0004-6361:20042488}

\bibitem[{Wang {et~al.}(2017)Wang, Li, Ma, \& Zhao}]{wang2017}
Wang, H.~H., Li, G.~Q., Ma, J.~H., \& Zhao, D. 2017, RSC Adv., 7, 44456, \dodoi{10.1039/C7RA07886B}

\bibitem[{{Wardle} \& {Kronberg}(1974)}]{wardle1974}
{Wardle}, J.~F.~C., \& {Kronberg}, P.~P. 1974, \apj, 194, 249, \dodoi{10.1086/153240}

\bibitem[{{Weingartner}(2006)}]{weingartner2006}
{Weingartner}, J.~C. 2006, \apj, 647, 390, \dodoi{10.1086/505342}

\bibitem[{{Weingartner} \& {Draine}(2001)}]{weingartner2001}
{Weingartner}, J.~C., \& {Draine}, B.~T. 2001, \apjs, 134, 263, \dodoi{10.1086/320852}

\bibitem[{{Whittet}(2003)}]{whittet2003}
{Whittet}, D. C.~B. 2003, Dust in the galactic environment - 2:nd ed. (Dust in the galactic environment Institute of Physics Publishing, 390 p.).
\newblock \url{http://adsabs.harvard.edu/cgi-bin/nph-bib_query?bibcode=2003QB791.W45......&db_key=AST}

\bibitem[{{Winters} {et~al.}(1994){Winters}, {Dominik}, \& {Sedlmayr}}]{winters1994}
{Winters}, J.~M., {Dominik}, C., \& {Sedlmayr}, E. 1994, \aap, 288, 255

\end{thebibliography}
\bibliographystyle{aasjournal}

\appendix

\section{Dipole model}\label{A1}

The magnetic field model is based on a functional form described by \citet{king2018} to predict the polarization in synthetic magnetohydronamical (MHD) simulations.  This model was then adapted to estimate the polarization from magnetically-aligned dust grains in a torus by \citet{lopezrodriguez2020}.  The implementation here has been simplified to model a rotationally-symmetric dipole field.

For the magnetic field configuration, we assume a rotationally-symmetric dipole of the form:
\begin{eqnarray}
  B_r &=& -2\,B_0\,\cos{\left(\phi+\psi\right)} \\
  B_\theta &=& -B_0\,\sin{\left(\phi+\psi\right)},
\end{eqnarray}
where $B_0$ is proportional to $r^{-2}$, $\phi$ is the azimuthal angle and $\psi$ is the pitch angle of the dipole. This represents a somewhat simplified dipole for two-dimensional computation.  In Cartesian coordinates, the field becomes:
\begin{eqnarray}
  B_x &=& \left(-B_\theta\,\sin{\phi}\right) + \left(B_r\,\cos{\phi}\right) \\
  B_y &=& \left(B_\theta\,\cos{\phi}\right) + \left(B_r\,\sin{\phi}\right) \\
  B_z &=& B_0(r).
\end{eqnarray}

When this dipole is viewed at an inclination angle $i$ and tilt $\Theta$ w.r.t.\ the plane of the sky, the observer's frame becomes:
\begin{eqnarray}
  B_{x^s} &=& B_x\,\cos{\Theta} + \left(B_y\,\cos{i}-B_z\,\sin{i}\right)\,\sin{\Theta} \\
  B_{y^s} &=& -B_x\,\sin{\Theta} + \left(B_y\,\cos{i}-B_z\,\sin{i}\right)\,\cos{\Theta} \\
  B_{z^s} &=& B_y\,\sin{i} + B_z\,\cos{i}.
\end{eqnarray}

Following \citet{lopezrodriguez2020}, we then express the Stokes parameters in terms of the local magnetic field $B=(B_x,B_y,B_z)$ in the optically-thin case $\tau_\lambda \ll1$, for each point in a map on the sky of RA, Dec ($\alpha, \delta$):
\begin{eqnarray}
  I(\alpha, \delta)&=&\int n\left(1-p_0\left(\frac{B_x^2+B_y^2}{B^2}-\frac{2}{3}\right)\right)\,ds \\
  Q(\alpha, \delta)&=&p_0\int n\left(\frac{B_y^2-B_x^2}{B^2}\right)\,ds \\
  U(\alpha, \delta)&=&p_0\int n\left(\frac{2B_xB_y}{B^2}\right)\,ds
\end{eqnarray}
where the $x$ and $y$ axes are in the plane of the sky, $z$ is along the line-of-sight, and $n$ describes the volume density.  In our simulations, ds is along the z-axis.  We implement $n$ as a two-dimensional Gaussian function to generate an arbitrary ``gas'' volume density.  $p_0$ relates dust grain cross-sections with magnetic alignment properties.  Because we are not, here, attempting to fit the polarization fraction, but only fit the position angles between data and model, and because dipoles are rotationally symmetric, the density distribution in the model is not of primary importance.   We note that, because the model does not include quantitative polarized radiative transfer calculations, we have limited our model fitting to the symmetric case with a zero inclination angle of the dipole axis, relative to the plane of the sky, e.g. a face-on view.

%
%

\end{document}